\documentclass[10pt,sigconf,letterpaper]{acmart}

\usepackage{adjustbox}
\usepackage{booktabs} 
\usepackage{subcaption}
\usepackage{enumerate}
\usepackage{cleveref}
\usepackage{array}
\usepackage{enumitem}
\usepackage{tabularx}
\usepackage{makecell}
\usepackage{multicol}
\usepackage{longtable}
\crefformat{footnote}{#2\footnotemark[#1]#3}

\usepackage{multirow}
\usepackage{listings}
\usepackage{wrapfig}

 \usepackage{xcolor}

\definecolor{dark green}{RGB}{94,145,40}

\newcommand{\drop}[1]{}
\newcommand{\ct}[1]{}
\newcommand{\ennote}[1]{}
\newcommand{\fsnote}[1]{}
\newcommand{\mlnote}[1]{}
\newcommand{\amnote}[1]{}
\newcommand{\mbnote}[1]{}
\newcommand{\ecnote}[1]{}
\newcommand{\isnote}[1]{}

\providecommand{\eg}{\emph{e.g.,} }

\newcommand{\enote}[2]{}

\copyrightyear{}
\acmYear{}
\setcopyright{none}
\acmConference[arXiv preprint]{arXiv preprint}{}{}
\acmBooktitle{}
\acmPrice{}
\acmDOI{}
\acmISBN{}

\renewcommand\footnotetextcopyrightpermission[1]{}
\lstdefinelanguage{json}{
    basicstyle=\normalfont\ttfamily,
    numbers=left,
    numberstyle=\scriptsize,
    stepnumber=1,
    numbersep=4pt,
    showstringspaces=false,
    breaklines=true,
    frame=lines,
    backgroundcolor=\color{background},
    literate=
     *{0}{{{\color{numb}0}}}{1}
      {1}{{{\color{numb}1}}}{1}
      {2}{{{\color{numb}2}}}{1}
      {3}{{{\color{numb}3}}}{1}
      {4}{{{\color{numb}4}}}{1}
      {5}{{{\color{numb}5}}}{1}
      {6}{{{\color{numb}6}}}{1}
      {7}{{{\color{numb}7}}}{1}
      {8}{{{\color{numb}8}}}{1}
      {9}{{{\color{numb}9}}}{1}
      {:}{{{\color{punct}{:}}}}{1}
      {,}{{{\color{punct}{,}}}}{1}
      {\{}{{{\color{delim}{\{}}}}{1}
      {\}}{{{\color{delim}{\}}}}}{1}
      {[}{{{\color{delim}{[}}}}{1}
      {]}{{{\color{delim}{]}}}}{1}
      {\ \ }{{\ }}1,
}

\lstdefinestyle{myCustomMatlabStyle}{
  keywords={
  	asname,
	cloud,
	dist,
	dst\_addr,
	dst\_asn,
	dst\_cc,
	dst\_continent,
	far\_side\_asn,
	hop\_cc\_best,
	hop\_continent\_best,
	hop\_rdns,
	hop\_rtt,
	near\_side_asn,
	probe\_ttl,
	src,
	src\_cc,
	src\_continent,
  },
alsoletter={\[ \]},
morekeywords={},
keywordstyle=\color{blue}\bfseries,
aboveskip=20pt,
belowskip=20pt,
identifierstyle=\color{black},
morecomment={[n][\color{purple}]{\#}{\^^M}},
numbers=left,
numberstyle=\color{black}\scriptsize,
rulecolor=\color{black},
stepnumber=1,
numbersep=8pt,
showstringspaces=false,
breaklines=true,
postbreak=\noindent,
breakautoindent=false,
frame=single,
backgroundcolor=\color{background},
}

\usepackage{pifont} 

\definecolor{darkgreen}{rgb}{0, 0.5, 0}

\definecolor{statecolor}{rgb}{0.85, 0.04, 0.1}
\definecolor{citycolor}{rgb}{0.1, 0.2, 0.8}
\definecolor{neighborhoodcolor}{rgb}{0.1, 0.6, 0.1}
\definecolor{streetcolor}{rgb}{0.6, 0.3, 0}
\definecolor{segmentcolor}{rgb}{0.5, 0, 0.5}
\definecolor{peeringcolor}{rgb}{0, 0.5, 0.5}
\definecolor{techcolor}{rgb}{0.8, 0, 0.8}

\usepackage{subcaption}

\usepackage{tikz}
\usetikzlibrary{mindmap,shadows.blur}
\usepackage{forest}

\makeatletter
\@ifpackageloaded{pxfonts}\@tempswafalse\@tempswatrue
\if@tempswa
  \DeclareFontFamily{U}{pxsymbols}{}
  \DeclareFontFamily{U}{pxAMSb}{}
  \DeclareSymbolFont{pxsymbols}{OMS}{pxsy}{m}{n}
  \SetSymbolFont{pxsymbols}{bold}{OMS}{pxsy}{bx}{n}
  \DeclareFontSubstitution{OMS}{pxsy}{m}{n}
  \DeclareSymbolFont{pxAMSb}{U}{pxsyb}{m}{n}
  \SetSymbolFont{pxAMSb}{bold}{U}{pxsyb}{bx}{n}
  \DeclareFontSubstitution{U}{pxsyb}{m}{n}
  \DeclareMathSymbol{\aleph}{\mathord}{pxsymbols}{64}
  \DeclareMathSymbol{\beth}{\mathord}{pxAMSb}{105}
  \DeclareMathSymbol{\gimel}{\mathord}{pxAMSb}{106}
  \DeclareMathSymbol{\daleth}{\mathord}{pxAMSb}{107}
\fi
\makeatother

\newif\iflinnaeusanon
\linnaeusanonfalse

\newcommand{\name}{\textit{Linnaeus}}

\let\oldquote\quote
\let\endoldquote\endquote
\renewenvironment{quote}[2][]
  {\if\relax\detokenize{#1}\relax
     \def\quoteauthor{#2}%
   \else
     \def\quoteauthor{#2~---~#1}%
   \fi
   \oldquote}
  {\par\nobreak\smallskip\hfill(\quoteauthor)%
   \endoldquote\addvspace{\bigskipamount}}

\usepackage{epigraph}

\definecolor{background}{RGB}{240,240,240}
\definecolor{boldblue}{RGB}{0,0,255}

\begin{document}
\title[\name{}]{\name{}: A Hierarchical, Multi-Label Framework for Autonomous System Classification}

\author{Marcos Piotto}
\affiliation{%
  \institution{Universidad de San Andr\'es}
  \country{Buenos Aires, Argentina}
}
\email{mpiotto@udesa.edu.ar}

\author{Ignacio Schuemer}
\affiliation{%
  \institution{Universidad de San Andr\'es}
  \country{Buenos Aires, Argentina}
}
\email{ischuemer@udesa.edu.ar}

\author{Santiago T. Torres}
\affiliation{%
  \institution{Universidad de San Andr\'es}
  \country{Buenos Aires, Argentina}
}
\email{storres@udesa.edu.ar}

\author{Mariano G. Beir\'o}
\orcid{0000-0002-5474-0309}
\affiliation{%
  \institution{Universidad de San Andr\'es}
  \country{Buenos Aires, Argentina}
}
\additionalaffiliation{%
  \institution{CONICET}
  \country{Buenos Aires, Argentina}
}
\email{mbeiro@udesa.edu.ar}

\author{Esteban Carisimo}
\orcid{0009-0002-1622-2305}
\affiliation{%
  \institution{Northwestern University}
  \country{Evanston, IL, USA}
}
\email{esteban.carisimo@northwestern.edu}

\author{Fabi\'an E. Bustamante}
\orcid{0000-0002-7659-1527}
\affiliation{%
  \institution{Northwestern University}
  \country{Evanston, IL, USA}
}
\email{fabianb@cs.northwestern.edu}

\renewcommand{\shortauthors}{M. Piotto et al.}

\begin{abstract}
Autonomous systems (ASes) play diverse roles in today's Internet, from community and research backbones to hyperscale content providers and submarine-cable operators. However, existing taxonomies based solely on network-level features fail to capture their semantic and operational heterogeneity. 

In this paper, we present \name{}, a hierarchical AS-classification framework that combines network-centric data (e.g., topology, BGP announcements) with rich non-network features and leverages domain-adapted large language models alongside traditional machine-learning techniques. \name{} provides a two-level taxonomy with 18 top-level and 38 second-level classes, supports multi-label assignments to reflect hybrid roles (e.g., research backbone + transit provider), and provides an end-to-end pipeline from data ingestion to label inference. On a manually annotated dataset of nearly 2,000 ASes, \name{} achieves an overall precision and recall of 0.83 and 0.76, respectively. 
We further demonstrate its practical value through case studies, highlighting \name{}'s potential to reveal both structural and semantic dimensions of Internet infrastructure.
\end{abstract}

\maketitle

\section{Introduction}


The Internet is a vast network of independently operated networks, or Autonomous Systems (ASes). Each AS represents a administrative domain that controls a group of IP prefixes and routing policies.

These ASes range widely in scale, ownership, and purpose. They include everything from community networks like NYC Mesh (AS395853) and individual ``hobbyist'' operators (e.g., AS200556, operated by Ethan Cady)~\cite{fiebig2024mauinmeowmeow, ripe2024personalasn}, to hypergiant content providers such as Meta (AS32934) and global transit backbones like Lumen (AS3356). Even small access networks, such as Ilha Turbo in Brazil (AS52803), and isolated deployments serving agricultural facilities (e.g., Ming Yi Tea Farm, AS17415) contribute to the Internet's decentralized character.

As the Internet has evolved beyond its academic roots, researchers have pursued a comprehensive taxonomy of its ASes to answer key questions about Internet structure, resilience, and control. Such classifications help illuminate how limited IPv4 address space is distributed -- for instance, highlighting the continued possession of underutilized /8 and /16 blocks by U.S. universities~\cite{dainotti2016lost} -- and reveal which networks serve as origins for large-scale attacks~\cite{antonakakis2017understanding} or exhibit persistent vulnerabilities to malware~\cite{moore2002code, shin2011large, moore2003inside}.

Moreover, AS-level classifications support analyses of security adoption (e.g., RPKI deployment across different AS types~\cite{testart2024identifying}), the role of submarine cable operators in cross-border connectivity, the robustness of government digital infrastructure, and the connectivity choices of critical services like power-grid operators. Ultimately, they reveal structural shifts and increasing centralization, as many formerly self-managed networks migrate to hyperscale cloud providers.

Despite decades of effort, a fine-grained and widely accepted taxonomy of ASes remains a challenge. 
Methods that draw exclusively on network-level data, such as topology snapshots, BGP relationships, or advertised address space, capture structural features yet overlook the semantic context that distinguishes, for example, a government backbone from a cloud platform or a hypergiant CDN from a community network. The Internet's vast heterogeneity, combined with the lack of consensus on classification criteria, compounds the difficulty. Only a handful of efforts have looked beyond traditional data sources, leveraging business intelligence platforms or crowd-sourced labels, but they still rely on sparse, often incomplete annotations that cannot reflect the full diversity of today's Internet. 
Recent breakthroughs in machine learning (ML), particularly domain-adapted large language models (LLMs) capable of combining heterogeneous evidence, open a promising path toward building a richer, more dynamic AS-classification framework.

In this paper, we introduce \name{}, a hierarchical framework for AS classification.
\name{} combines state-of-the-art LLMs with traditional ML techniques and leverages both network-centric data and rich, non-network features to generate a two-level taxonomy comprising 18 top-level and 38 second-level classes. 
In contrast to prior work that assigns each AS a single role, we adopt a multi-label strategy, allowing \name{} to capture organizations with multiple operational domains --- for example, Internet2 (AS11164) functions simultaneously as a research backbone and a transit provider. 
This design reconciles the network-oriented and business-oriented paradigms that have historically remained separate.
Rather than enforcing mutually exclusive categories, \name{} captures the overlapping operational roles that modern ASes routinely play.

We demonstrate that \name{} is an effective approach.
Evaluated on a manually curated dataset of $\approx$ 2,000 ASes, \name{} achieves precision and recall of 0.83 and 0.76, respectively. 
We demonstrate its practical value by classifying the $\approx$120,000 allocated ASNs and expanding its value through two case studies: (i) the footprint of government-run networks in the Internet's network, (ii) the scale of educational and research networks.

Our key contributions are:

\begin{itemize}
\item We introduce an open-source framework for autonomous-system classification.\footnote{\iflinnaeusanon The full codebase will be released upon paper acceptance.\else Available at \url{https://github.com/NU-AquaLab/linnaeus}.\fi}
\item By combining complementary machine-learning paradigms, our approach captures multiple operational dimensions of Internet networks more effectively than prior work.
\item We publish a corpus of nearly 2,000 manually annotated ASes --- the largest gold-standard dataset for AS classification to date --- and characterize the resulting landscape across all $\approx$120k allocated ASNs.
\end{itemize}

This paper does not raise any ethical considerations.
\section{Background}

Despite decades of work, classifying ASes remains a persistent challenge. Since the Internet's transition to commercial use and today's ecosystem of hyperscalers, personal ASes, and critical infrastructure networks, researchers have struggled to develop taxonomies that reflect both the functional and organizational diversity of ASes. Early efforts focused on small, manually curated datasets and simple network-centric features. Later approaches introduced richer attributes, such as PeeringDB metadata and customer-cone metrics, but still fell short of capturing ASes' real-world roles. In this section, we first review key prior efforts (\S\ref{sec:background:relwork}), highlighting their contributions and limitations. We then outline how advances in AI and LLMs offer new opportunities to rethink AS classification (\S\ref{sec:background:ml}), enabling richer, more context-aware frameworks like \name{}.

\subsection{Related Work}\label{sec:background:relwork}

Several efforts have explored streamlined and automated AS taxonomies over the last two decades. Dimitropoulos et al.~\cite{dimitropoulos2006revealing} introduced one of the earliest AS taxonomies. 
By manually inspecting 1,200 ASes, this method proposed six categories,\footnote{Large ISPs, small ISPs, customer networks, universities, Internet exchange points, and network information centers.} omitting both transit providers and the then-emerging content-delivery networks (CDNs), and a classifying approach solely based on 3 network attributes: organization records, announced IP prefixes, and inferred AS relationships

The growing importance of CDNs, with a handful of providers generating the majority of the network traffic, motivates Bottger et al.~\cite{bottger2018looking} to develop an automated identification technique. Leveraging the rich data available in PeeringDB~\cite{peeringdb2025}, including port capacity, geographic footprint, and traffic profile, the authors applied threshold-based heuristics to flag content-provider networks that dominate global traffic.

Also building on PeeringDB data, CAIDA~\cite{caida2025} maintained an AS-classification dataset~\cite{caida2015asclassification}. 
The dataset results from combining PeeringDB attributes with the output of a supervised classifier that ingests various features: AS-level degree metrics (customer, provider, peer), customer-cone size, advertised address space, the number of Alexa Top-1M domains served by the AS, and the fraction of its prefixes observed as active by the UCSD Network Telescope. 
Despite this rich input, the resulting taxonomy included only three classes: Transit/Access, Content, and Enterprise networks.

More recently, Ziv et al.~\cite{ziv2021asdb} introduced ASdb, which shifts the focus from classifying ASes to classifying the organizations that own them. ASdb utilizes features of various natures: business intelligence platforms (Dun \& Bradstreet, Crunchbase, ZoomInfo, Clearbit), network datasets (PeeringDB, IPinfo), and automated website profiling (Zvelo). 
Using the North American Industry Classification System (NAICS)~\cite{naics2025}, which encompasses 17 top-level industries and 95 sub-sectors, ASdb combines rule-based heuristics, ML website classifiers, and crowd-sourced validation via Amazon Mechanical Turk, to classify network owners. Despite its rich feature set, the NAICS taxonomy may not offer a complete picture of the diversity of networks operating on the Interne and the gold-standard training set of only 150 randomly selected AS owners risks bias in the Internet's highly imbalanced ecosystem.

Industry actors have also entered the field, for example, IPinfo's Tags service~\cite{ipinfo2025tags} and BGP.Tools~\cite{bgptools2025tags} assigns descriptive labels to ASes to support coarse-grained classification at scale. 

Beyond such commercial efforts, numerous research papers introduce ad-hoc labels tailored to a single category or study, typically derived through manual curation~\cite{kumar:govt_choices, carisimo:state_owned_ases, richter2018advancing, ager2011web, bajpai2017vantage}.

\subsection{New Paths for AI/ML Methods}\label{sec:background:ml}

Recent breakthroughs in ML, especially the rise of transformer-based foundation models, have significantly enhanced the reach of ML. 
LLMs such as GPT-4~\cite{openai2023gpt4} and Llama 3~\cite{touvron2023llama} are trained on multi-trillion-token corpora and are now available either as open weights or public APIs. 
Their off-the-shelf generality eliminates the need for highly specialized architectures and significantly reduces reliance on carefully curated, domain-specific datasets.

Zero-shot and few-shot prompting further magnify these advantages~\cite{brown2020language}. 
With only a handful or no labeled examples, LLMs can tackle classification, named entity recognition (NER), and other information-extraction (IE) tasks that once demanded large annotation campaigns~\cite{brown2020language}. 
These models have also absorbed vast swaths of the public web, carrying rich latent representations of Internet actors --- from network operators and IXPs to cloud and CDN providers. 
This embedded knowledge enables the identification and categorization of networking artefacts with minimal additional labeling effort.

Borges~\cite{borges:imc} and The Aleph~\cite{aleph:conext} are recent examples of domain-adapted LLMs, which demonstrate how these general capabilities translate to networking tasks. 
Borges employs information-extraction techniques to infer AS-to-organization mappings from sibling relationships embedded in PeeringDB records. 
The Aleph generates regular expressions that reveal geo-hints in PTR records, guided by a few prompt-and-example pairs. 
Together, these systems highlight the potential of LLMs to automate labor-intensive data-curation tasks in Internet measurement.

Earlier AS-classification efforts, developed before the advent of LLMs, relied primarily on network-centric features and lacked the ability to incorporate non-network contextual data. While some studies explored machine learning and early NLP techniques, they fell short in extracting rich semantic information, such as identifying whether a network is operated by a government agency, university, or utility provider. This limitation stemmed not only from the absence of models capable of semantic inference, but also from the resulting reliance on structural network features alone.

A promising research direction we explore in this work is to enrich AS classification by integrating traditional network-centric features (e.g., customer cone size) with external semantic signals, such as organization names, descriptions, and websites. This could be enabled by combining multimodal datasets with both LLM-based zero- or few-shot inference and more conventional machine learning approaches.


\section{Categories}

This section introduces \name{}, a classification scheme for ASes that reflects how their operators engage with the Internet ecosystem. Unlike prior efforts that focused narrowly on either network function or business role, \name{} bridges both perspectives through a two-level hierarchy: 18 top-level categories and 38 more specific subclasses. This structure offers significantly finer granularity than existing taxonomies, and adopts a multi-label design that favors interpretability over strict mutual exclusivity.

Throughout this paper, we use the term \emph{taxonomy} to denote a structured classification scheme with explicitly defined categories, hierarchical parent-child relationships, and documented decision criteria for label assignment --- as opposed to a flat tagging system, which assigns labels without hierarchical structure or mutual constraints. \name{}'s two-level hierarchy, with 18 top-level categories and 38 sub-classes linked by conditional membership rules, satisfies this definition. The multi-label design does not contradict this: labels are not arbitrary tags but operationally grounded categories whose co-occurrence is constrained by the hierarchy (e.g., \emph{executive} is only valid within \emph{government}). The rest of this section defines each category and describes the types of organizations they represent.

\begin{figure*}[ht!]
	\centering
	\includegraphics[width=0.75\textwidth]{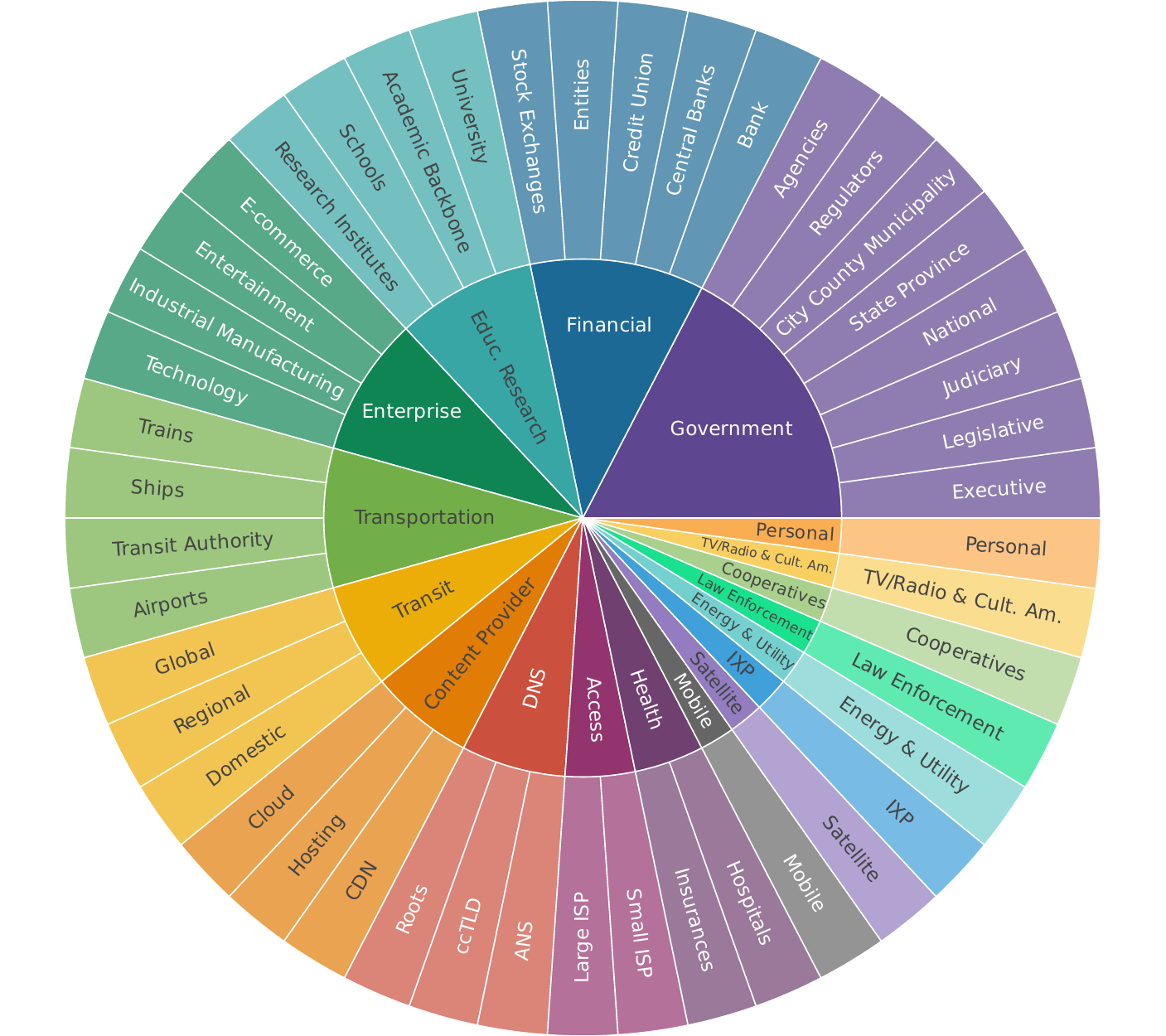}
	\caption{\name{}'s taxonomy: top-level and sub-level categories}
	\label{fig:taxonomy}
\end{figure*}

\noindent{\textbf{Access Networks}} provide last-mile connectivity, linking end-users to the global Internet: \textit{small access networks} serve up to a few thousand subscribers in a narrow area, while \textit{large access networks} serve millions and often span entire countries.
We distinguish the two tiers using CAIDA customer-cone size: ASes above the 80th percentile of the cone-size distribution for access-tagged networks are classified as large; all others as small. This threshold was selected by inspecting the empirical distribution, which exhibits a natural break around the 80th percentile separating a small number of dominant national providers from the long tail of community, SOHO, and boutique ISPs --- consistent with known consolidation patterns in national broadband markets.

\noindent{\textbf{Transit Providers}} announce more IP prefixes than they originate, generating revenue by carrying third-party traffic. We classify them into three geographic tiers --- \textit{Domestic}, \textit{Regional}, and \textit{Global} --- based on RIR registration data and IX presence.

\noindent{\textbf{Non-Terrestrial Networks}} include \textit{satellite operators} (GEO, MEO, LEO) and \textit{mobile providers}.\footnote{We treat mobile virtual network operators (MVNOs) as mobile where they operate their own ASN. Table~\ref{tab:mvno-asn} in \S~\ref{sec:appendix:mvno} presents examples.}

\noindent{\textbf{DNS Infrastructure}} encompasses networks whose primary function is DNS operation, labeled by role: \textit{Root DNS}, \textit{(cc)TLD registry}, or \textit{Authoritative DNS hosting}.

\noindent{\textbf{IXP}} includes ASNs belonging to Internet exchange points that facilitate traffic exchange between networks.

\noindent{\textbf{Content Providers}} serve primarily outbound content traffic; we subdivide them into \textit{CDN}, \textit{Cloud}, and \textit{Hosting}, excluding transits that offer CDN services (e.g., Level3).

\noindent{\textbf{Educational and Research}} captures ASes operated by educational and research institutions: \textit{universities}, \textit{schools}, \textit{research institutes}, and \textit{academic backbones}.

\noindent{\textbf{Government}} ASes are classified along three orthogonal axes --- \textit{jurisdiction} (national, state/province, city/municipality), \textit{branch} (executive, legislative, judicial), and \textit{role} (regulator, agency) --- to capture the diverse forms of oversight across public networks.

\noindent{\textbf{Law Enforcement}} labels networks operated by public-safety agencies such as police departments and coast-guard services.

\noindent{\textbf{Enterprise}} includes industry-specific subcategories: \textit{technology}, \textit{manufacturing/industrial}, \textit{e-commerce}, and \textit{entertainment}.

\noindent{\textbf{Energy \& Utilities}} covers providers of essential services: \textit{electricity}, \textit{gas}, \textit{water}, and \textit{energy/oil}.

\noindent{\textbf{Finance}} encompasses the financial sector: \textit{commercial banks}, \textit{central banks}, \textit{credit unions}, and \textit{stock exchanges}.

\noindent{\textbf{Health}} distinguishes \textit{hospitals} from \textit{insurance} providers.

\noindent{\textbf{Cooperatives}} captures member-owned networks such as Mid-Rivers Telephone Cooperative (AS11961).

\noindent{\textbf{TV, Radio \& Cultural Amenities}} covers networks dedicated to cultural broadcasting or heritage services: \textit{TV channels}, \textit{radio stations}, and \textit{libraries/museums}.

\noindent{\textbf{Transportation}} covers transport-service ASes: \textit{airlines/airports}, \textit{railways}, \textit{shipping companies}, and \textit{transit authorities}.

\noindent{\textbf{Personal}} ASes belong to individuals operating networks for hobbyist or experimental purposes~\cite{ripe2024personalasn,fiebig2024mauinmeowmeow}.

The resulting taxonomy yields operationally meaningful labels that distinguish economic roles while accommodating the Internet's heterogeneity. Detailed per-subcategory definitions with representative examples appear in \S~\ref{sec:appendix:taxonomy_details}.

\section{Features}\label{sec:features}

The set of features that \name{} adopts as classification inputs blends \textit{network-centric attributes} (\S\ref{sec:features:network})---such as topological properties and routing patterns---with \textit{semantics-rich data} (\S\ref{sec:features:nonnetwork}) drawn from organizational names and website content. The following subsections define each feature and explain our rationale for its inclusion.

\subsection{Network-Centric Attributes}\label{sec:features:network}

\name{} characterizes each AS through three classes of network-centric attributes: (i) access and transit role, (ii) geographic reach, and (iii) traffic characteristics such as routing asymmetry and aggregate prefix space. To assess {\bf access roles}, \name{} examines advertised address space and ``eyeball'' population --- the number of end-user hosts behind the AS. Large prefix footprints often indicate large access networks, but exceptions (e.g., legacy university allocations~\cite{dainotti2016lost} or NATed environments~\cite{richter2016multi, sherry2012making}) motivate the inclusion of application-layer measurements of end-user density. To capture {\bf transit roles}, \name{} compares originated address space against customer-cone space, using large discrepancies to flag ASes that mainly carry others' traffic; customer and provider degree and cone size further refine this view. {\bf Geographic footprint} is evaluated both directly --- by counting peering facilities and aggregating their distribution across cities, countries, and continents --- and indirectly --- by measuring diversity within the customer cone, i.e., distinct downstream-country counts distinguishing regional ISPs from global backbones. Finally, {\bf traffic profiles} rely on public proxies since operators guard traffic data: \name{} considers (i) traffic asymmetry, (ii) self-declared capacity tiers (e.g., 10–100 Gbps, 1 Tbps+), and (iii) aggregate IXP port capacity, which together provide an operational sketch without exposing proprietary metrics.

\subsection{Semantics-Rich Attributes}\label{sec:features:nonnetwork}

\name{} augments structural attributes with semantics-rich signals that offer insight into a network's real-world role. Specifically, it extracts (i) the AS name, (ii) the legal name of the operating organization, (iii) registered countries for both the AS and its owner, and (iv) hyperlinks published on the operator's website. When processed by a fine-tuned LLM, these contextual features help differentiate, for example, a government-operated backbone from a commercial CDN, even when their topologies look alike. These linguistic cues are essential for capturing functionally meaningful distinctions across the Internet ecosystem.

\section{Model}

In the following paragraphs, we describe the architecture of \name{}'s model (\S\ref{sec:model:overview}), followed by a detailed description of its component (\S\ref{sec:model:llm}~--~\S\ref{sec:model:sc}).

\subsection{Architecture: Building Blocks and Design Choices}\label{sec:model:overview}

\begin{figure*}[ht!]
	\centering
	\includegraphics[width=0.95\textwidth]{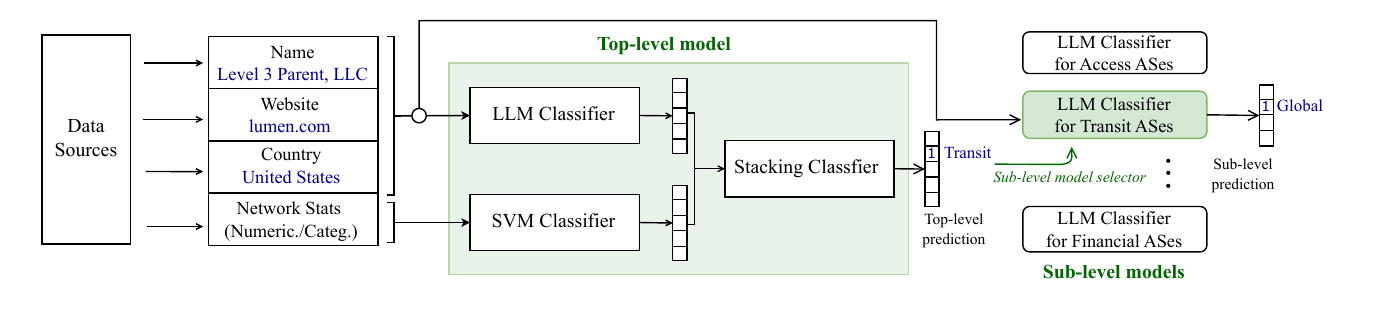}
	\caption{Model architecture for the multilabel tag prediction task.}
	\label{fig:architecture}
\end{figure*}

\name{}'s pipeline, shown in Figure~\ref{fig:architecture} operates in two tiers. 
A top-level classifier first assigns a set of tags to each AS matching the categories defined in Section 3. Then, for each assigned tag, dedicated sub-classifiers refine that decision to produce category-specific second-level tags.
All stages draw on semantics-rich features, while the top-level model additionally incorporates network-centric attributes.

The top-level classifier adopts a hybrid architecture (green-shaded block in Fig.~\ref{fig:architecture}) with three modular stages: (1) {\bf LLM stage} --- a domain-adapted large-language model that extracts semantic signals from business records, websites, and other unstructured sources; (2) {\bf SVM stage} --- a support-vector machine that learns from traditional network features such as network degree, customer cone, and PeeringDB attributes; and (3) {\bf Stacking meta-classifier} --- a lightweight ensemble that combines the outputs of the LLM and SVM to produce the final tags.

The sub-classification layer relies exclusively on LLMs that ingest the top-level prediction together with the semantics-rich features. 
After the top-level model assigns a broad tag, each sub-classifier is utilized only for the sub-tags valid within the category pointed out by that tag.
For instance, the sub-tag \emph{executive} is considered only if the first-level classifier has already tagged the AS as \emph{government}. 
This cascading design progressively narrows the candidate tags set, minimizing misclassifications and enforcing the mutually exclusive constraints of our taxonomy.

\subsection{The LLM stage: Extracting Semantic Information}\label{sec:model:llm}
Each LLM stage (top-level and subclassifiers) consumes a high-level feature prompt formed by concatenating (i) AS and organization names, (ii) their registered countries, and (iii) the operator's public website. This context is passed to a fine-tuned instance of OpenAI's GPT-4o-mini, which returns a probability distribution over candidate tags plus a short structured rationale that we later embed for stacking. We fine-tune separate models at each hierarchy level (e.g., government vs.\ commercial; executive vs.\ judicial), and at inference use instruction-style prompts including the taxonomy schema, a few in-context examples, and decoding constraints to exclude invalid labels (see \S\ref{sec:results:training}). Few-shot prompting complements fine-tuning, and we employ chain-of-thought prompting to elicit stepwise reasoning. For I/O, the model emits a fixed JSON schema (e.g., {\tt\{label:\dots, confidence:0.\dots\}}) to ensure deterministic parsing. To improve throughput, we use mini-batch prompting (up to 10 AS records per request), reducing API overhead and improving in-context contrast, and we fix {\tt temperature}=0 and {\tt top\_p}=1 for reproducible outputs.

\subsection{The SVM stage: Extracting Network Information}\label{sec:model:svm}

To complement the semantic cues from the LLM, the top-level model incorporates a Support Vector Machine (SVM) that processes structured, tabular features beyond the language model's scope. We use an RBF-kernel SVM for its ability to model nonlinear decision boundaries in high-dimensional space. Its input vector includes (1) {\bf Topology}: total, provider-, and customer-specific AS degree; cone sizes; and counts of prefixes and IP addresses; (2) {\bf Geography}: country and continent as one-hot vectors; and (3) {\bf Capacity and user statistics}: aggregate IXP port capacity and estimated subscriber counts. In preliminary evaluations, the SVM consistently outperformed tree-based models (e.g., XGBoost), so we adopt it as the default structured-data classifier. Missing values are imputed using {\tt KNNImputer} (or {\tt SimpleImputer} with zero fill when sparse), and all numeric features are standardized via {\tt StandardScaler} to zero mean and unit variance to balance kernel sensitivity. The resulting probability scores are then passed to the stacking meta-classifier for integration with LLM outputs.

\subsection{The Stacking Classifier: Reconciling Both Views}\label{sec:model:sc}

The final stage of the top-level model uses a stacking classifier,\footnote{Implemented using Scikit-Learn's \texttt{StackingClassifier}.} which integrates semantic signals from the fine-tuned LLM with numerical features learned by the structured-data SVM. Each base estimator outputs class-probability vectors that serve as inputs to a meta-learner, implemented as a linear-kernel SVM. This meta-learning layer fuses unstructured and structured evidence into a unified decision space, automatically weighting each modality by its predictive utility.


\section{Datasets}\label{sec:dataset}

In this section, we describe the datasets we use to implement \name{}.
We enumerate the external sources from which we derive our feature set (\S\ref{sec:dataset:datasources}) and describe our process for constructing the gold-standard corpus used to train and evaluate the model (\S\ref{sec:dataset:annotated}).

\subsection{Data Sources}\label{sec:dataset:datasources}

To supply the features in \S\ref{sec:features}, \name{} merges five complementary datasets: CAIDA AS-Rank~\cite{caida2025asrank}, APNIC Eyeball~\cite{apnic_eyeballs}, PeeringDB~\cite{peeringdb2025}, IPinfo~\cite{ipinfo2025}, and RDAP~\cite{ietf2015rdap}. Together, these provide network-topology statistics, traffic indicators, and semantically rich metadata (Table~\ref{tab:dataset}).  
From {\bf AS-Rank} (Dec 1, 2024 snapshot, 118,519 ASNs), we extract access metrics (originated address space), transit metrics (customer-cone size, degree, total address space), geography (customer-cone countries, regions, continents), and identity fields (country, AS and organization names). {\bf APNIC Eyeball} data (Jan 1, 2025, 30,919 ASes) estimate end-user populations behind NAT using Google Ad traffic~\cite{eyeballs1}, validated in~\cite{apnic:imc24}. {\bf PeeringDB} (Jan 3, 2025, 31,548 ASes) contributes operator-reported names, websites, traffic tiers, asymmetry, geographic scope, and IXP presence (used to infer port capacity and diversity); coverage varies by attribute due to voluntary reporting. {\bf IPinfo} (Apr 23, 2025) supplements missing URLs for 83,078 ASes; for ASes present in AS-Rank but missing here, we manually scrape sites, yielding complete name/URL coverage for 119,748 ASes. Finally, {\bf RDAP} fills residual gaps in organizational and contact metadata.  
Most ASes appearing only in AS-Rank are small access networks or dormant allocations. Because many hypergiants (e.g., Google, Meta, Microsoft) require current PeeringDB entries for public peering~\cite{microsoft:peering:policy, google:peering:policy, facebook:peering:policy}, missing records generally indicate limited operational scope.

\begin{table}[httb]
	\scriptsize
	\setlength{\tabcolsep}{5pt}
	\begin{tabular*}{\columnwidth}{@{\extracolsep{\fill}} l r  l r @{}}
		\toprule
		\multicolumn{2}{c}{\textbf{PeeringDB (2025-01-03)}} &
		\multicolumn{2}{c}{\textbf{Other sources}} \\
		\cmidrule(lr){1-2}\cmidrule(lr){3-4}
		Metric & \# Records & Metric & \# Records \\
		\midrule
		Total network records & 31,548 & IPinfo total ASNs (2025-04-23) & 83,078 \\
		Website & 26,895 & \quad Website field present & 76,512 \\
		Network-type & 22,531 & CAIDA AS-Rank ASNs (2024-12-01) & 118,519 \\
		Geo-scope & 31,164 & \quad Cones inferred & 77,495 \\
		Peering-ratio & 30,860 & AS-Pop total records (2025-01-01) & 30,919 \\
		Facility ({\tt netfac}) & 12,783 & & \\
		IXP LAN ({\tt netixlan}) & 16,345 & & \\
		\bottomrule
	\end{tabular*}
	\caption{Coverage statistics: detailed PeeringDB fields (left) and summary counts from IPinfo, CAIDA AS-Rank, and AS-Pop (right).}
	\label{tab:dataset}
\end{table}

\subsection{Annotated Dataset}\label{sec:dataset:annotated}

To build a gold-standard corpus for supervised training and evaluation, we manually labeled a diverse subset of ASes. 
We began by drawing a uniform random sample from CAIDA's AS-RANK export and assigning each AS a taxonomy tag. 
Since the Internet is heavily skewed toward small access networks, pure random sampling produced too few examples of more uncommon categories.
To correct this imbalance, we (i) added all ASes appearing in the AS-Rank top-100 to enhance coverage of large transit providers, (ii) queried Hurricane Electric's BGP portal with category-specific keywords --- in English, Spanish, Portuguese, Italian, French, German and Russian\footnote{Our language choice is based on content's language popularity worldwide and across regions~\cite{isoc2023languages}.} ---  to obtain additional candidates, and (iii) incorporated artefacts from prior work (e.g., requested a list of federal-run networks~\cite{kumar:govt_choices}). 
We continued sampling until every category contained at least 22 examples, yielding a final dataset of 1,870 manually annotated ASes. A detailed per-category breakdown appears in \S~\ref{sec:appendix:annotated:class_stats}.

In Appendix~\ref{sec:appendix:annotated:availability}, we also report the percentage of per-feature data available across all annotated AS categories, visualized as a heatmap that highlights coverage of each feature group (see Fig.~\ref{fig:heatmap}).
\section{Model Training and Evaluation}
\label{sec:results}

In this section, we detail how we trained \name{} and computed  its performance metrics. 
We describe details of the training procedure (\S\ref{sec:results:training}), and report the overall accuracy as well as first- and second-level results (\S\ref{sec:results:acc}).

\subsection{Training and Evaluation using Cross-Validation}
\label{sec:results:training}

\begin{figure*}[htbp]
	\centering
	\begin{subfigure}[b]{0.95\textwidth}
		\centering
		\includegraphics[width=\textwidth]{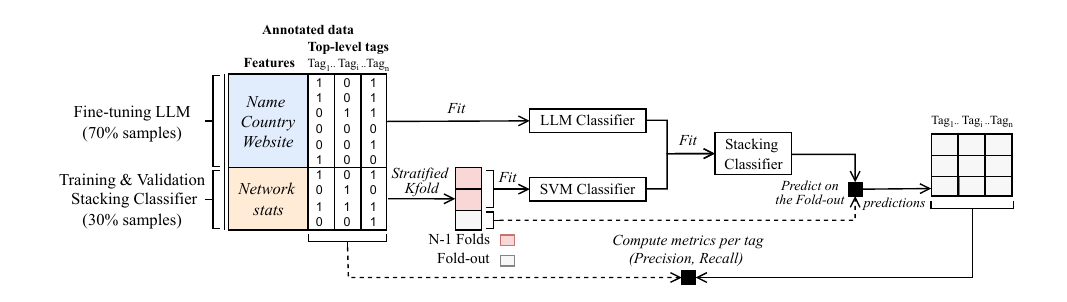}
		\caption{Top-level model training and evaluation pipeline}
		\label{fig:top_level_pipeline}
	\end{subfigure}
	\vspace{1em}
	\begin{subfigure}[b]{0.90\textwidth}
		\centering
		\includegraphics[width=\textwidth]{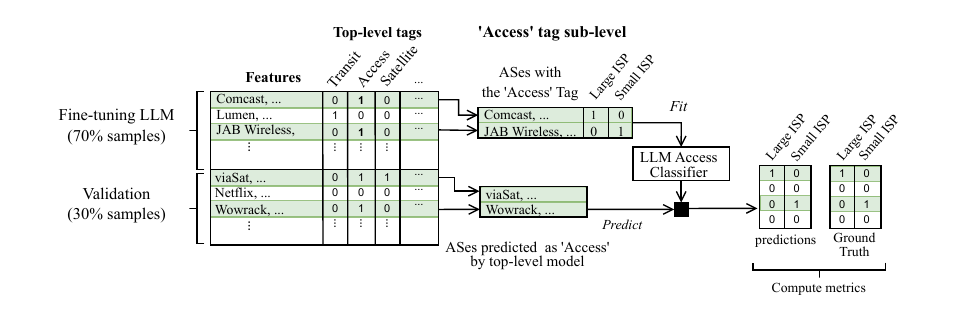}
		\caption{Sub-level model training and evaluation pipeline}
		\label{fig:sub_level_pipeline}
	\end{subfigure}
	\caption{Pipeline for training and evaluation of the top-level model (Fig.~\ref{fig:top_level_pipeline}) and each sub-level model (Fig.~\ref{fig:sub_level_pipeline}).}
	\label{fig:pipeline_training}
\end{figure*}

Our training pipeline (Fig.~\ref{fig:pipeline_training}) required several practical implementation decisions, primarily driven by data availability and quality rather than the model's conceptual design.  The following paragraphs outline those decisions and the criteria that guided them.

\textbf{Data prioritization.} Because several sources overlap (see \S\ref{sec:dataset:datasources}) --- e.g., organization names, AS names, website URLs, and country codes --- we rank them to give the LLM the cleanest inputs: IPinfo serves as the primary source for its normalized, human-readable labels; PeeringDB is the fallback when IPinfo lacks a field; and RDAP is consulted only as a last resort since its registry-style entries often lack structured encodings. This hierarchy is applied consistently when populating website and country attributes to ensure the richest semantic context.  

\textbf{LLM fine-tuning.} The top-level and sub-models are fine-tuned independently using semantic attributes (organization name, country, website) from $70\%$ of the annotated sample (Fig.~\ref{fig:pipeline_training}). The top-level model predicts high-level categories, while each sub-model assigns zero or more sub-tags to ASes within its respective category.

\textbf{Performance evaluation.} We evaluate the full pipeline via 3-fold cross-validation: two folds for training and one for testing, using the remaining $30\%$ of annotated ASes not seen during fine-tuning. Nested CV handles hyperparameter tuning, and the best model is selected via $\mathrm{F}_\beta$ with $\beta=0.5$ to favor precision. The stacking classifier is implemented via \texttt{MultiOutputClassifier} in \texttt{scikit-learn}, training one model per tag. 

Figure~\ref{fig:top_level_pipeline} shows how network statistics feed into the SVM classifier, which in turn is combined with LLM outputs in the final stacking layer.  

\textbf{Managing unbalanced, multi-label data.} The Internet's inherent skew --- whether in connectivity~\cite{yook2002modeling}, traffic~\cite{labovitz2010internet}, or organizational diversity --- produces an imbalanced multi-label corpus dominated by a few categories. To maintain proportional representation across folds, we apply iterative stratification~\cite{sechidis2011stratification, szymanski2017network}, which preserves tag-combination frequencies in each split. Table~\ref{tab:split_comparison} compares train-test splits for a subclass of {\em Transit} and illustrates why this method is essential for balanced partitions.\footnote{We use the {\tt skmultilearn} library: \url{http://scikit.ml/api/skmultilearn.model_selection.iterative_stratification.html}.}  

\textbf{Cost and runtime.} Fine-tuning all models costs roughly \textbf{\$5}, and inference over $\sim$120k ASes adds about \textbf{\$25}. Each fine-tuning run takes about 15 minutes, with sub-level models trained in parallel (three at a time), for a total of about one hour. Inference over all ASes requires $\sim$12 hours, while fitting the stacking classifier takes only a few minutes and is negligible in comparison.

\begin{table}[htbp]
	\footnotesize
	\centering
	\caption{Example comparison of data-splitting methods by subclass for the class {\em Transit}.}
	\label{tab:split_comparison}
	\begin{tabular}{l
			*{4}{c}
			*{4}{c}}
		\toprule
		& \multicolumn{4}{c}{No stratification}
		& \multicolumn{4}{c}{Iterative} \\
		\cmidrule(lr){2-5}\cmidrule(lr){6-9}
		Class        & Orig & Train & Val & Val \% 
		& Orig & Train & Val & Val \% \\
		\midrule
		Domestic     & 24   & 13    & 11  & 45.83\% 
		& 24 & 17    & 7   & 29.17\% \\
		Global       & 42   & 32    & 10  & 23.81\% 
		& 42 & 29    & 13  & 30.95\% \\
		Regional     & 10   & 8     & 2   & 20.00\% 
		& 10 & 7     & 3   & 30.00\% \\
		\bottomrule
	\end{tabular}
\end{table}

\subsection{Model Accuracy}\label{sec:results:acc}

Next, we evaluate \name{} on the gold-standard corpus, reporting performance for both the top-level classifier and each sub-classifier.

Beyond per-class scores, we use three global summaries: (1) Macro metrics of precision/recall/F1 score, (2) average per-label accuracy and (3) subset accuracy. Macro metrics (macro-precision/recall/F1) compute each metric per label and then take the unweighted mean across them, so rare and common labels count equally~\cite{sklearn_precision_score_docs}. Average per-label accuracy computes one accuracy per label and then averages them. Subset accuracy is the strictest metric for multi-label classification, measuring the fraction of samples whose predicted label set exactly matches the ground-truth label set for them.

\begin{table*}[ht]
  \scriptsize
  \centering
  \caption{Per-tag metrics (Prec: Precision, Rec: Recall, Acc: Accuracy) for the top-level model. The average per-label accuracy is 0.97, the macro precision is 0.83, the macro recall is 0.76, and the subset accuracy (exact match) is 0.66.}
  \label{tab:general_tags_metrics_sorted_by_precision}
  \begin{tabular*}{\textwidth}{@{\extracolsep{\fill}} l c c c  l c c c @{}}
    \toprule
    \textbf{Tag} & \textbf{Prec.} & \textbf{Rec.} & \textbf{Acc.} 
                & \textbf{Tag} & \textbf{Prec.} & \textbf{Rec.} & \textbf{Acc.} \\
    \midrule
    \textbf{Personal}               & 0.88 & 0.78 & 0.99 
                                   & \textbf{DNS}                  & 0.87 & 0.79 & 0.98 \\
    \textbf{Health}                 & 0.94 & 0.94 & 1.00 
                                   & \textbf{Transit}              & 0.80 & 0.69 & 0.98 \\
    \textbf{Transportation}         & 0.87 & 0.84 & 0.98 
                                   & \textbf{Finance}              & 0.90 & 0.90 & 0.98 \\
    \textbf{IXP}                    & 0.95 & 0.84 & 0.98 
                                   & \textbf{Educational \& Research}  & 0.92 & 0.83 & 0.97 \\
    \textbf{Access}                 & 0.85 & 0.72 & 0.93 
                                   & \textbf{Government}           & 0.83 & 0.82 & 0.90 \\
    \textbf{Cooperatives}           & 0.85 & 0.85 & 0.99 
                                   & \textbf{Satellite}            & 0.89 & 0.67 & 0.99 \\
    \textbf{Law Enforcement}        & 0.86 & 0.75 & 0.99 
                                   & \textbf{Mobile}               & 0.73 & 0.77 & 0.97 \\
    \textbf{Energy \& Utility}      & 0.68 & 0.68 & 0.98 
                                   & \textbf{Tv, Radio \& Cult. Amens.} & 0.75 & 0.69 & 0.99 \\
    \textbf{Content Providers}      & 0.79 & 0.63 & 0.95
                                   & \textbf{Enterprise}           & 0.66 & 0.49 & 0.93 \\
    \bottomrule
  \end{tabular*}
\end{table*}

Table~\ref{tab:general_tags_metrics_sorted_by_precision} reports \name{}'s top-level performance. 
The model reaches an average per-label accuracy of 0.97, a subset accuracy of 0.66, with macro-averaged precision and recall of 0.83 and 0.76, respectively. 
Given that our corpus is sharply skewed --- true negatives are abundant --- precision and recall are the most meaningful indicators of effectiveness. 
Per-class precision ranges from 0.95 for IXP ASes to 0.66 for enterprise networks. 
The latter reflects the semantic mismatch between enterprise and content-provider roles: many companies self-declare as content providers, and, in principle, any host running an open service could qualify as one. 
These ambiguities remain challenging even for a semantics-aware classifier.

\begin{table*}[httb]
  \scriptsize
  \centering
  \caption{Per-tag metrics (Prec: Precision, Rec: Recall, Acc: Accuracy) for all sub-level models. The average per-label accuracy is 0.98, the macro precision is 0.79, the macro recall is 0.71, and the subset accuracy (exact match) is 0.59.}
  \label{tab:per_label_metrics_hierarchical_compact}
  \begin{tabular*}{\textwidth}{@{\extracolsep{\fill}} lccc lccc}
    \toprule
    \textbf{Tag} & \textbf{Prec.} & \textbf{Rec.} & \textbf{Acc.} &
    \textbf{Tag} & \textbf{Prec.} & \textbf{Rec.} & \textbf{Acc.} \\
    \midrule
    \textbf{Access}               &       &       &       & \textbf{DNS}                  &       &       &       \\
    {Small ISP}   & 0.681 & 0.701 & 0.929 & {Roots}         & 1.000 & 1.000 & 1.000 \\
    {Large ISP}   & 0.605 & 0.821 & 0.966 & {(cc)TLDs}      & 0.778 & 0.700 & 0.992 \\
    \textbf{Transit}             &       &       &       & {Auth Nameservers} & 1.000 & 0.125 & 0.988 \\
    {Global}      & 0.381 & 0.800 & 0.975 & \textbf{Energy \& Utility}     & 0.684 & 0.684 & 0.980 \\
    {Regional}    & 0.500 & 0.182 & 0.981 & \textbf{Enterprise}            &       &       &       \\
    {Domestic}    & 0.571 & 0.400 & 0.985 & {E-commerce}    & 1.000 & 0.429 & 0.993 \\
    \textbf{Mobile}              & 0.727 & 0.774 & 0.973 & {Entertainment} & 0.556 & 0.500 & 0.985 \\
    \textbf{Satellite}           & 0.889 & 0.667 & 0.992 & {Industrial \& Manuf.} & 1.000 & 0.429 & 0.993 \\
    \textbf{Content Provider}    &       &       &       & {Technology}   & 0.467 & 0.206 & 0.941 \\
    {Cloud}       & 0.875 & 0.700 & 0.993 & \textbf{Financial}             &       &       &       \\
    {Hosting}     & 0.538 & 0.412 & 0.946 & {Bank}          & 0.926 & 0.926 & 0.993 \\
    {CDN}         & 0.800 & 0.727 & 0.992 & {Central Banks} & 0.857 & 0.857 & 0.997 \\
    \textbf{Educational Research}&       &       &       & {Credit Union}  & 1.000 & 1.000 & 1.000 \\
    {University}  & 0.762 & 0.889 & 0.988 & {Entities}      & 0.667 & 0.600 & 0.988 \\
    {Academic BB} & 0.667 & 0.800 & 0.990 & {Stock Exchanges} & 0.900 & 0.900 & 0.997 \\
    {Schools}     & 0.941 & 0.667 & 0.985 & \textbf{Law Enforcement}       & 0.857 & 0.750 & 0.990 \\
    {Research Inst.} & 0.800 & 0.800 & 0.986 & \textbf{Health}                &       &       &       \\
    \textbf{Government}          &       &       &       & {Insurances}    & 1.000 & 1.000 & 1.000 \\
    {Executive}   & 0.653 & 0.807 & 0.875 & {Hospitals}     & 0.889 & 1.000 & 0.998 \\
    {Legislative} & 1.000 & 0.750 & 0.997 & \textbf{Cooperatives}          & 0.846 & 0.846 & 0.993 \\
    {Judiciary}   & 1.000 & 0.737 & 0.992 & \textbf{TV, Radio \& Cult. Amens.} & 0.750 & 0.692 & 0.988 \\
    {National}    & 0.788 & 0.775 & 0.912 & \textbf{Transportation}        &       &       &       \\
    {State/Prov.} & 0.800 & 0.774 & 0.978 & {Trains}        & 1.000 & 1.000 & 1.000 \\
    {City/Municip.} & 0.636 & 0.667 & 0.975 & {Ships}         & 1.000 & 0.857 & 0.998 \\
    {Regulators}  & 0.400 & 0.600 & 0.978 & {Transit Authority} & 0.833 & 0.833 & 0.997 \\
    {Agencies}    & 0.808 & 0.656 & 0.973 & {Airports}      & 0.833 & 0.769 & 0.992 \\
    \textbf{IXP}                 & 0.947 & 0.837 & 0.985 & \textbf{Personal}              & 0.875 & 0.778 & 0.995 \\
    \bottomrule
  \end{tabular*}
\end{table*}

Table~\ref{tab:per_label_metrics_hierarchical_compact} reports per-subcategory accuracy, precision, and recall. 
\name{} achieves an average per-label accuracy of 0.98, a macro precision of 0.79, a macro recall of 0.71, and a subset accuracy of 0.59. 
The model performs remarkably well in semantically dense domains, accurately discriminating among government branches, education and research institutions, and the various types of financial organizations and enterprises, highlighting its capacity for fine-grained Internet taxonomy. 
In contrast, distinctions among transit providers lag behind. 
This may be due to the increasingly blurred boundaries between domestic, regional, and global transit. The widespread use of remote peering and the adoption of IRUs to access hubs such as NYIIX and DE-CIX could make it challenging to clearly distinguish among these categories.

\subsection{Cross-Taxonomy Analysis: Why a Networking-Specific Taxonomy?}\label{sec:results:asdb}

Economic classification systems such as NAICS and ISIC were never designed to 
distinguish transit providers from access networks, or academic backbones from 
universities --- roles that are central to Internet topology research. 
\name{}'s {\em bring-your-own-taxonomy} (BYOT) interface allows us to test 
this structural mismatch directly, by training \name{} on these external 
schemas and measuring what is lost when networking-specific distinctions have 
no label to land on.


\paragraph{Taxonomies}
NAICS --- the North American Industry Classification System --- is a hierarchical, six-digit schema used by U.S., Canadian, and Mexican statistical agencies to classify establishments; it is periodically revised and widely adopted for economic reporting.
ASdb proposes {\em NAICSlite} as a compact mapping of NAICS tailored to networking use cases, collapsing the full hierarchy into a concise label set while retaining broad sectoral coverage.
ISIC --- the International Standard Industrial Classification maintained by the U.N. --- is the global standard for classifying economic activities and is used in official statistics.
The two taxonomies are closely related, and official correspondence tables provide sector-level mappings between NAICS and ISIC.
Crucially, neither taxonomy was designed to capture roles unique to organizations that operate ASes --- distinctions such as transit vs.\ access, academic backbones vs.\ universities, or IXP vs.\ content provider have no counterparts in economic classification systems.

\paragraph{Methodology}
Using the 2,000 ASNs in our manually annotated dataset, we extracted the corresponding ASdb labels from the January 2024 dataset and trained \name{} to predict \emph{NAICSlite} (and, via published correspondences, \emph{ISIC}) categories.
To enable a fair comparison, we fixed the dataset size and used a 70/30 train-test split, preserving label proportions with iterative stratification.
The model architecture and features were unchanged; only the target label set varied.

\paragraph{Step 1: Diagnosing ASdb label quality (not a generalization benchmark)}
As a diagnostic first step — not a generalization benchmark — we train \name{} directly on ASdb's original annotations for our 2,000 ASes.
Performance is limited --- macro precision = 0.57, macro recall = 0.48, macro F1 = 0.50 --- as detailed in Table~\ref{tab:dry_run} (Appendix~\ref{sec:appendix:asdb_training}).
A targeted error analysis reveals that this degradation stems largely from label noise in the existing ASdb annotations rather than from \name{}'s inability to learn the taxonomy.

To characterize the noise, we manually cross-referenced all 2,000 ASdb-labeled records against AS names, PeeringDB entries, and organizational websites. This review revealed systematic misclassifications throughout: for example, NVIDIA (AS398037) is labeled as a museum and Cloudflare (AS13335) as government. Such errors are not isolated incidents — they appear across multiple categories and are consistent with ASdb's limited gold-standard set of 150 ASes.
We therefore proceed to a full manual curation of all 2,000 records to build 
a cleaner gold standard (released with our repository) before evaluating 
generalization.

\begin{table*}[ht]
  \scriptsize
  \centering
  \setlength{\tabcolsep}{2.5pt}
  \caption{Per-tag metrics (Prec: Precision, Rec: Recall, Acc: Accuracy) for the \name{} model (fine-tuned) to the ASdb (NAICSlite) taxonomy using our annotations as Ground Truth. Average per-label Accuracy = 0.98, Macro Precision = 0.85, Macro Recall = 0.76, Macro F1 = 0.78, Subset Accuracy = 0.80.}
  \label{tab:linnaeus_model_on_asdb_taxonomy}
  \begin{tabular*}{\textwidth}{@{\extracolsep{\fill}} l c c c  l c c c @{}}
    \toprule
    \textbf{Tag} & \textbf{Prec.} & \textbf{Rec.} & \textbf{Acc.}
                & \textbf{Tag} & \textbf{Prec.} & \textbf{Rec.} & \textbf{Acc.} \\
    \midrule
    \textbf{Other}                         & 1.00 & 1.00 & 1.00 
                                   & \textbf{Retail \& E-commerce}               & 1.00 & 0.86 & 1.00 \\
    \textbf{Agriculture, Mining \& Refining} & 1.00 & 0.50 & 1.00 
                                   & \textbf{Freight \& Postal}                  & 1.00 & 0.44 & 0.98 \\
    \textbf{IT \& Software}                & 0.94 & 0.93 & 0.95 
                                   & \textbf{Finance \& Insurance}               & 0.92 & 0.95 & 0.99 \\
    \textbf{Education \& Research}         & 0.91 & 0.87 & 0.97 
                                   & \textbf{Travel \& Lodging}                  & 0.89 & 0.89 & 0.99 \\
    \textbf{Manufacturing}                 & 0.88 & 0.78 & 0.99 
                                   & \textbf{Government \& Public Admin}         & 0.87 & 0.86 & 0.92 \\
    \textbf{NGOs \& Community}             & 0.87 & 0.59 & 0.98 
                                   & \textbf{Media \& Publishing}                & 0.85 & 0.50 & 0.98 \\
    \textbf{Health Care}                   & 0.83 & 0.94 & 0.99 
                                   & \textbf{Utilities (non-ISP)}                & 0.82 & 0.69 & 0.99 \\
    \textbf{Museums \& Entertainment}      & 0.60 & 1.00 & 1.00
                                   & \textbf{Services (General)}                 & 0.25 & 0.33 & 0.99 \\
    \bottomrule
  \end{tabular*}
\end{table*}

\paragraph{Step 2: Generalization on NAICSlite with curated ground truth}
Having established that ASdb's original labels are too noisy to serve as a fair benchmark, we re-annotated all 2,000 records manually and evaluate \name{} on the same NAICSlite taxonomy with this clean ground truth. Using our curated gold-standard annotations as ground truth, we evaluate \name{} on the ASdb (\textit{NAICSlite}) taxonomy.
\name{} attains a macro precision = 0.85, macro recall = 0.76, macro F1 = 0.78, and subset accuracy = 0.80; per-category results appear in Table~\ref{tab:linnaeus_model_on_asdb_taxonomy}.
On the average per-label accuracy --- the metric used by ASdb to evaluate their own approach --- \name{} achieves 0.98 versus 0.93 for ASdb, on a substantially larger annotated corpus (2,000 vs.\ 150 in ASdb).

\paragraph{\name{} on the UN's ISIC taxonomy}
In parallel with our NAICSlite labeling effort, we manually assigned ISIC codes to the same 2,000 ASes.
\name{} achieves an average per-label accuracy = 0.99, Macro Precision = 0.85, Macro Recall = 0.84, Macro F1 = 0.83, Exact-Match Accuracy = 0.83, as detailed in Table~\ref{tab:isic_results} in \S~\ref{sec:appendix:isic}.

"\textbf{Generalization summary.} When evaluated against curated ground truth, \name{} achieves F1 $\geq$ 0.78 on NAICSlite and F1 = 0.83 on ISIC — confirming that the architecture generalizes across label systems. The low scores in Table~\ref{tab
	:dry_run} (F1 = 0.50) reflect ASdb label noise, not model failure.
	
\paragraph{What this tells us about taxonomy design}
The high performance under both NAICSlite (F1 = 0.78) and ISIC (F1 = 0.83) when trained on \emph{curated} labels — versus F1 = 0.50 when trained on ASdb's noisy original labels — confirms that \name{}'s architecture generalizes, and that label quality is the binding constraint.
However, this finding also reveals a deeper structural issue: general-purpose economic taxonomies such as NAICSlite and ISIC were not designed to capture the operational roles unique to network operators.
Categories like transit provider, academic backbone, IXP, or satellite operator have no counterpart in standard industrial classification.
Any classifier trained on these schemas must collapse operationally distinct network types into the same economic bucket, reducing discrimination precisely where Internet topology research needs it most.
This structural mismatch --- not \name{}'s architecture --- is what limits cross-taxonomy generalization, and it is the primary motivation for \name{}'s networking-specific taxonomy.

\section{Exploring the AS landscape}

We now analyze \name{}'s classification of the entire allocated AS namespace. We begin by characterizing the distribution of top-level tags (\S\ref{sec:analysis:all}), and then focus on two semantically rich domains: government ASes and education-and-research networks (\S\ref{sec:analysis:govt}).

\subsection{Analysis of top-level tags}\label{sec:analysis:all}

We start by examining the category distribution assigned by \name{} across all allocated ASNs. To assess labeling accuracy, we manually reviewed 50 randomly sampled ASes per class and found the model's predictions to be consistently accurate.

\begin{figure*}[httb]
  \centering
  \includegraphics[width=0.9\textwidth]{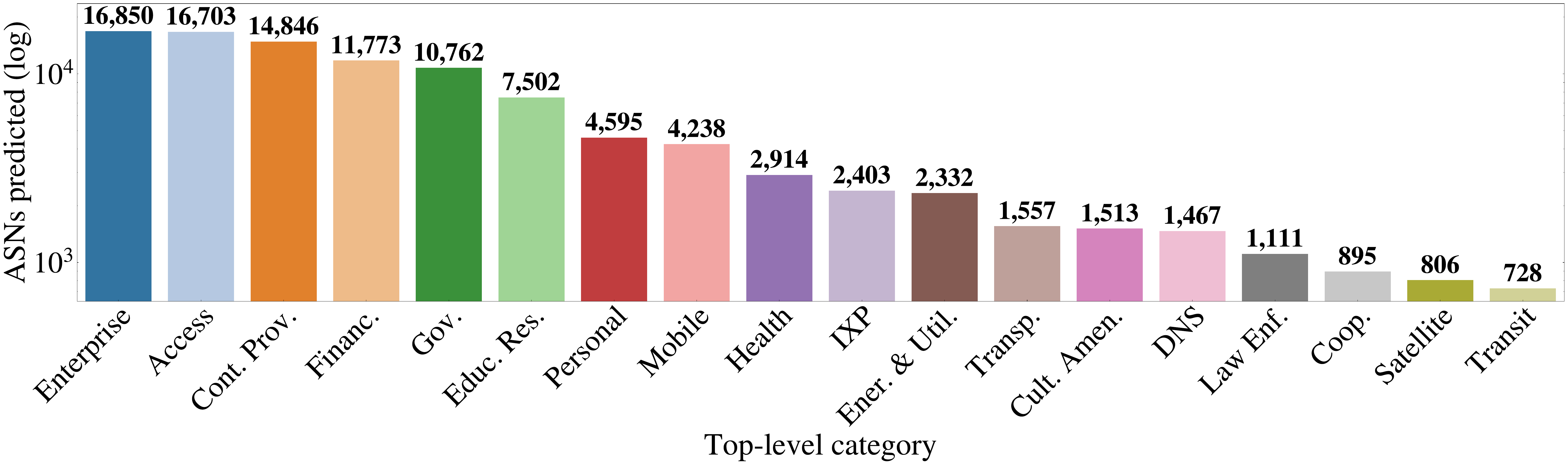}
  \caption{Top-level distribution for the 118,695 tagged ASes.}
  \label{fig:topLevelAllPredictions}
\end{figure*}

Figure \ref{fig:topLevelAllPredictions} shows the distribution of \name{}'s top-level tags for the 118,695 ASNs in our dataset. 
Most ASes fall into the {\em Enterprise} or {\em Access} categories, followed by {\em Content Providers}. The latter is particularly notable: many ASes self-identify as content providers while merely offering basic hosting services, without competing with CDNs. 

As \name{} supports multi-label assignments, we also examined tag combinations. 
In total, 81,838 ASNs received a single tag, 10,032 received two tags, and 333 received three tags. 
A residual 26,469 ASNs remained unlabeled at the top level, either because they lacked sufficient data or fell outside our current taxonomy; these ASNs were not assigned any second-level classes, as the sub-classifiers are conditioned on a valid top-level prediction.

\subsection{Exploring Government and Education Sub-Classifications}\label{sec:analysis:govt}

Last we examine the sub-categories within the {\em Government}, {\em Educational}, and {\em Research} domains.

\begin{figure*}[htbp]
    \centering
    \begin{subfigure}[b]{0.48\textwidth}
        \centering
        \includegraphics[width=\textwidth]{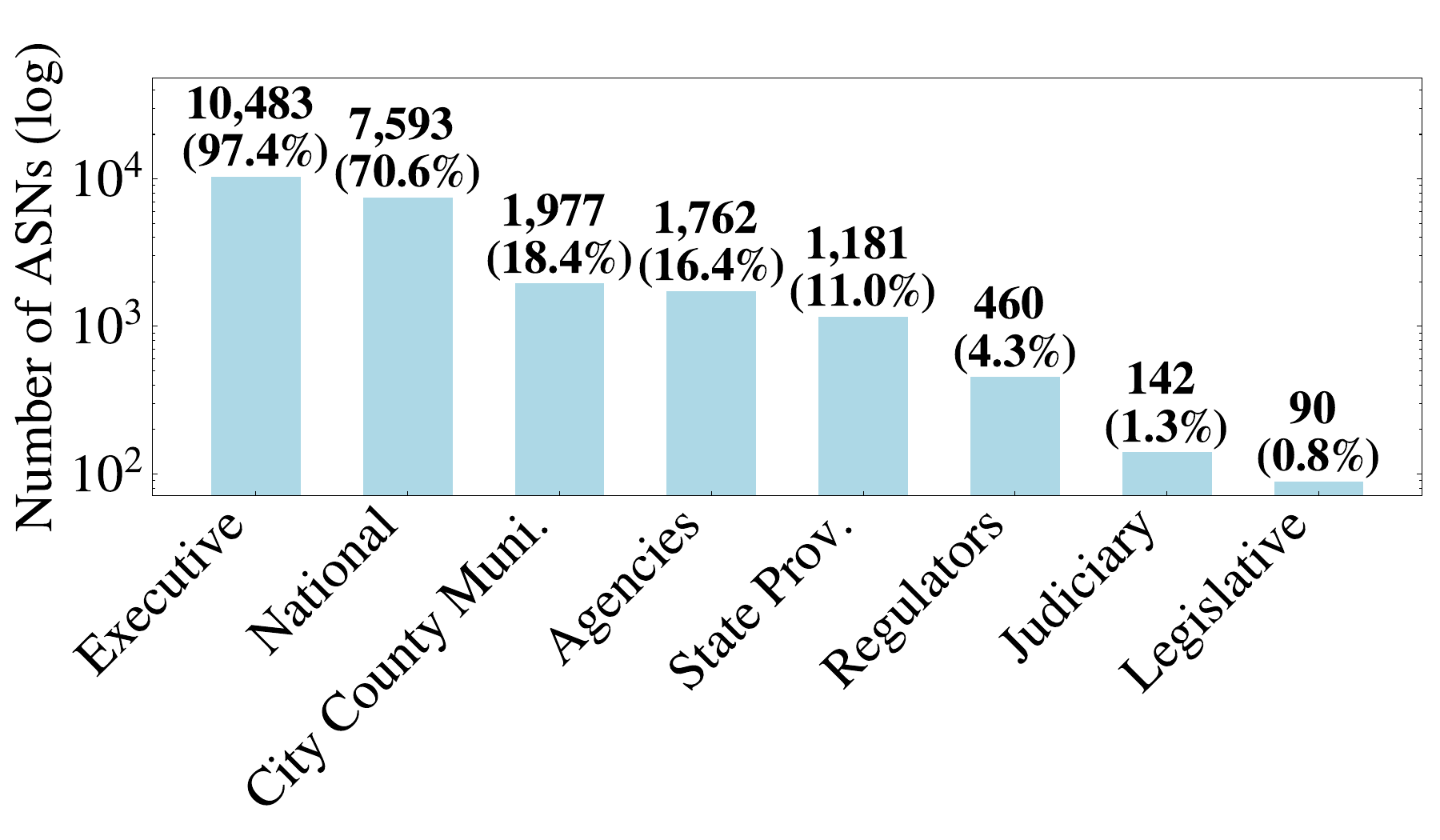}
        \caption{Government sub-level tags.}
        \label{fig:sub:govt}
    \end{subfigure}~\begin{subfigure}[b]{0.480\textwidth}
        \centering
        \includegraphics[width=\textwidth]{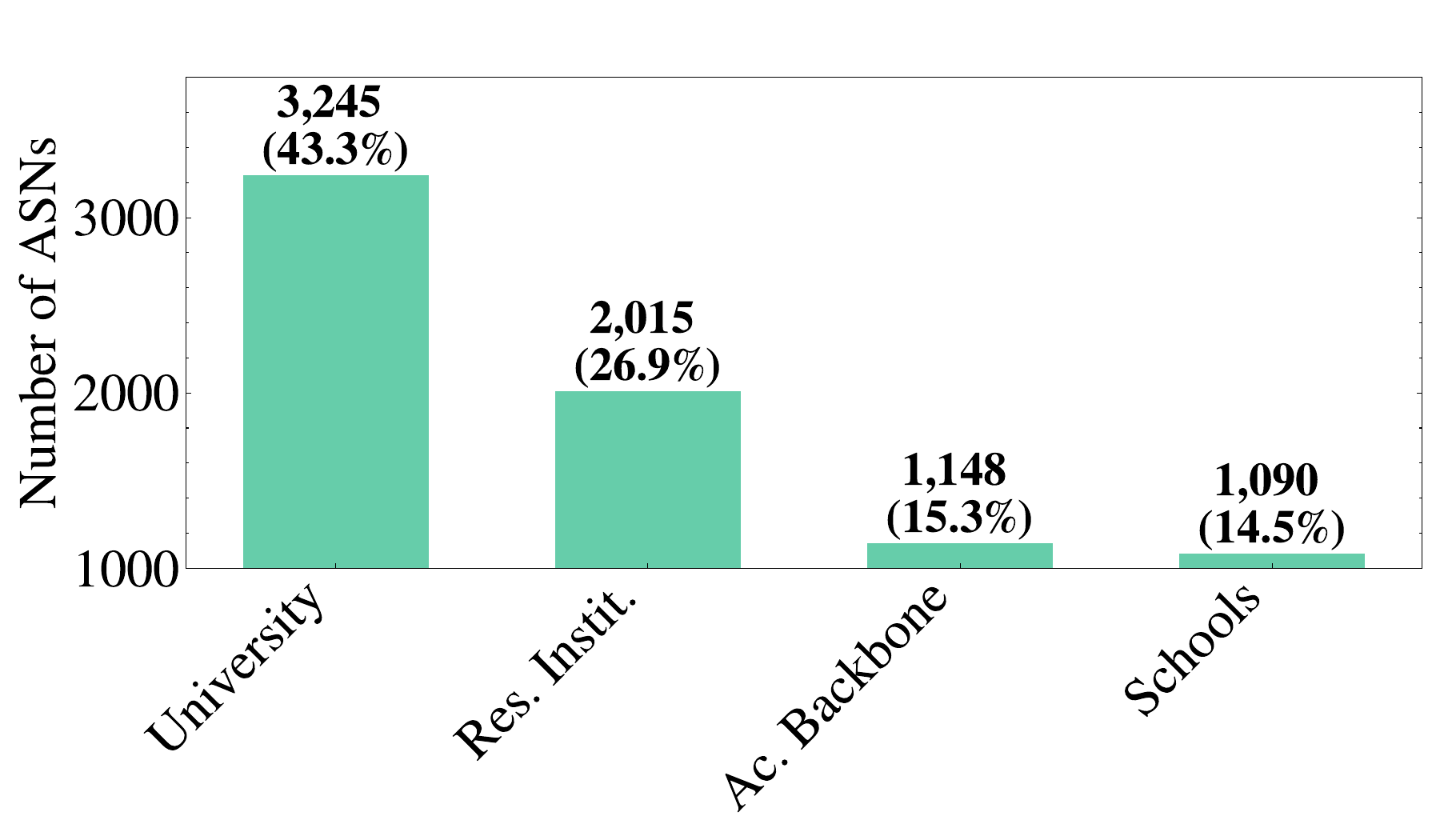}
        \caption{Education and Research sub-level tags.}
        \label{fig:sub:edu}
    \end{subfigure}
    \caption{Sub-level tags for government and educational, and research networks.}
    \label{fig:sub}
\end{figure*}


Figure~\ref{fig:sub} analyzes the government and education-and-research sectors. Among the 10,762 government-tagged ASes, nearly 97\% fall under the executive branch. Jurisdiction skews heavily national: 70.6\% operate at the federal level, with state/provincial and municipal entities accounting for 18.4\% and 11.0\%, respectively. \name{} identifies 5,402 ASes as both {\em National} and {\em Executive} (e.g., AS1977 -- Navy Network Information Center, US), and 1,672 that also carry the `agency' label (e.g., AS41336 -- Financial Agency, Croatia).

In the education-and-research sector, {\em Universities} lead with 43.4\% of ASes -- unsurprising given the Internet's academic roots -- followed by {\em Research institutes} (26.9\%, e.g., AS50507 -- European Academy of Bolzano, Italy) and {\em Academic Backbones} (15.3\%, e.g., AS2231 -- Renater, France). Notably, 722 ASes carry the combined tags {\em Research Institute}, {\em Executive}, and {\em National} (e.g., AS132158 -- Australian Synchrotron), reflecting large, state-funded institutions with their own network infrastructure.
\section{Related Work}\label{sec:relwork}

\textbf{AS classification and taxonomy.}
The challenge of classifying ASes has attracted sustained attention. Dimitropoulos et al.~\cite{dimitropoulos2006revealing} pioneered the field with a six-category taxonomy derived from WHOIS records and network-level attributes, using a small manually annotated corpus of 1,200 ASes. CAIDA~\cite{caida2015asclassification} later produced a three-class taxonomy (Transit/Access, Content, Enterprise) by combining PeeringDB features with supervised classifiers trained on customer-cone metrics, prefix counts, and telescope observations. Both efforts demonstrate that network-centric features alone produce coarse taxonomies that fail to capture the organizational diversity of the modern Internet.

\textbf{PeeringDB-based approaches.}
The proliferation of PeeringDB~\cite{peeringdb2025} as a registration database for networks worldwide has enabled richer feature engineering. Bottger et al.~\cite{bottger2018looking} leverage PeeringDB port capacity and traffic profiles to automatically identify hypergiants --- networks that dominate global traffic. ASdb~\cite{ziv2021asdb} extends this direction by mapping AS operators to the NAICS economic taxonomy using a combination of business intelligence data (Dun \& Bradstreet, Crunchbase), automated website classifiers, and crowd-sourced annotation via Amazon Mechanical Turk. While ASdb scales to more categories than prior work, its gold-standard training set of 150 ASes introduces sampling bias in the Internet's heavily skewed ecosystem. \name{} builds on these foundations with a substantially larger annotated corpus (2,000 ASes) and a taxonomy designed from the ground up to reflect Internet operational roles rather than economic sectors.

\textbf{IXP detection and peering topology.}
Internet exchange points occupy a structurally distinct role in the AS ecosystem. Several tools identify IXPs from route-server prefixes, RIPE RIS data, and PeeringDB membership records~\cite{augustin2009ixp}. Richter et al.~\cite{richter2018advancing} and related work on peering link inference highlight how IXP presence shapes transit relationships. \name{} dedicates a first-level category to IXPs and correctly identifies them at high precision (0.95), reflecting the importance of accurate IXP labeling for topology studies.

\textbf{LLMs for Internet measurement.}
The application of large language models to networking tasks is nascent but growing. Borges~\cite{borges:imc} uses information-extraction techniques to infer AS-to-organization mappings from PeeringDB sibling relationships. The Aleph~\cite{aleph:conext} applies LLMs to PTR-record geo-hint extraction via few-shot prompting. Both systems demonstrate that LLMs can automate data-curation tasks that previously required manual effort. \name{} extends this line of work to the classification problem, combining fine-tuned GPT-4o-mini for semantic label inference with an SVM trained on network-centric features and stacking both into a unified classifier.

\textbf{Government and critical-infrastructure AS studies.}
A distinct body of work targets specific AS categories at depth. Carisimo et al.~\cite{carisimo:state_owned_ases} investigate state-owned enterprise ASes, and Kumar et al.~\cite{kumar:govt_choices} study government network policies across jurisdictions. Bajpai et al.~\cite{bajpai2017vantage} examine research and education networks. \name{} subsumes these bespoke studies into a unified, multi-label framework and provides classifiers for government branches, academic backbones, law enforcement, and other categories previously studied in isolation.

\textbf{Alternative taxonomies and cross-taxonomy transfer.}
Industry actors including IPinfo~\cite{ipinfo2025tags} and BGP.Tools~\cite{bgptools2025tags} assign coarse descriptive tags to ASes for operational use. \name{} differs in two respects: it provides fine-grained two-level labels and supports \emph{bring-your-own-taxonomy} (BYOT) transfer to external schemas such as NAICSlite and ISIC (\S\ref{sec:results:asdb}). This cross-taxonomy capability distinguishes \name{} from prior AS classifiers, which are tightly coupled to their training label sets.

\section{Discussion}
\label{sec:discussion}

Our study has several limitations, summarized below.

\textbf{Coverage and biases.} While PeeringDB offers coverage, it underrepresents small and niche networks, and inactive ASes remain invisible to our network-centric dataset. Consequently, our taxonomy should be viewed as a conservative lower bound on the diversity of the Internet.  

\textbf{Expanding the feature space.} Future work could incorporate richer signals --- (i) active measurements that fingerprint server software and TLS stacks, (ii) DNS records (PTR, CNAME) revealing hosting relationships, and (iii) large-scale scans (e.g., Censys, Shodan) exposing service footprints. Proprietary scan datasets were considered but excluded due to access restrictions; this study relies exclusively on public data sources.  

\textbf{Taxonomy granularity and corner cases.} Any global taxonomy must balance specificity and generality. Some entities (e.g., AS400801, the Bi-State Development Agency spanning two states) defy strict categorization, and semantically meaningful but sparsely represented labels (e.g., political parties, bus operators) lack enough samples for dedicated models --- suggesting future manual curation of the ``long tail''.  

\textbf{Reliance on self-declared attributes.} Several fields --- cloud status, network type, peering policy --- are self-reported in PeeringDB. While generally reliable~\cite{lodhi2014using}, such declarations sometimes diverge from our semantic definitions (e.g., ``cloud providers'' offering only colocation). These discrepancies introduce noise; resolving them will require external validation (e.g., API offering).

\textbf{Dormant ASNs.} Nearly one-third of allocated ASNs appear inactive~\cite{anghel2024driving}. Low entry costs~\cite{arin2018fees} and legacy allocation policies~\cite{nemmi2021parallel} have led to stockpiles of ``personal'' ASNs. Dormant allocations inflate the search space while emitting no operational signals, complicating automated classification. Unless reclaim-and-recycle policies gain traction~\cite{ripe2025quarantine}, this silent tail will persist.  

\textbf{Taxonomy scope and external comparisons.} Our cross-taxonomy analysis (\S\ref{sec:results:asdb}) demonstrates that general-purpose economic classifications (NAICSlite, ISIC) cannot substitute for a networking-specific taxonomy. When trained on curated labels, \name{} achieves strong performance on both schemas (F1 $\geq$ 0.78), but the schemas themselves conflate operationally distinct network types --- transit and access providers, for example, both fall under ``IT \& Software'' in NAICS. This is not a failure of our classifier; it is structural evidence that the Internet's operational diversity requires a purpose-built taxonomy. Extending \name{} to cover additional networking-specific dimensions (\eg submarine cable operators, IXP route servers) remains an important direction for future work.

\textbf{LLM hallucinations and interpretability.} Although LLMs perform well in few-shot classification~\cite{brown2020language, chae2025large}, they may hallucinate by assigning incorrect categories~\cite{xu2024hallucination}. We quantify such errors through out-of-sample precision and recall, and mitigate them via chain-of-thought prompting, fine-tuning, and an SVM complement. Still, the interpretability of LLM outputs remains limited~\cite{divya2024comparing}, marking a clear direction for future work.

\section{Conclusions}

We presented \name{}, a new taxonomy and classification pipeline for Autonomous Systems that combines network-centric attributes with semantics-rich contextual data. By leveraging recent advances in LLms and pairing them with established topological features, \name{} overcomes the limitations of prior AS classification efforts that relied on narrow, often siloed indicators. Our two-level taxonomy captures the diversity of roles ASes play across the global Internet ecosystem, from dominant transit providers and content platforms to government agencies, cooperatives, and hobbyist networks. We show that combining linguistic signals with structural features yields fine-grained classifications that are both interpretable and operationally relevant.

As the Internet's complexity accelerates, understanding the function and diversity of its constituent networks becomes increasingly critical. We hope \name{} serves as a foundation for future work in Internet measurement, resilience analysis, and infrastructure accountability.

\bibliographystyle{ACM-Reference-Format}
\bibliography{base,refs} 

@string{sigcomm-ccr = "ACM SIGCOMM CCR"}

@string{sigcomm-ccr = "ACM SIGCOMM Computer Communication Review"}

@string{pnas = "Proc. of the National Academy of Sciences (PNAS)"}

@string{TPRC= "Telecommunication Policy Research Conference"}

@string{NSDI="Proc NSDI"}

@string{NSDI="Proceedings of the USENIX Symposium on Networked Systems Design and Implementation"}

@string{imc = "Proc. of IMC"}

@string{sigcomm = "Proc. of ACM SIGCOMM"}

@string{nsdi = "Proc. of USENIX NSDI"}

@string{pam = "Proc. of PAM"}

@string{conext = {Proc. of CoNEXT}}

@inproceedings{dimitropoulos2006revealing,
  title={Revealing the autonomous system taxonomy: The machine learning approach},
  author={Dimitropoulos, Xenofontas and Krioukov, Dmitri and Riley, George and others},
  booktitle=pam,
  year={2006}
}

@inproceedings{ziv2021asdb,
  title={ASdb: a system for classifying owners of autonomous systems},
  author={Ziv, Maya and Izhikevich, Liz and Ruth, Kimberly and Izhikevich, Katherine and Durumeric, Zakir},
  booktitle=imc,
  year={2021}
}

@article{bottger2018looking,
  title={Looking for hypergiants in peeringDB},
  author={B{\"o}ttger, Timm and Cuadrado, Felix and Uhlig, Steve},
  journal=sigcomm-ccr,
  year={2018},
}

@misc{caida2015asclassification,
  author       = {{Center for Applied Internet Data Analysis (CAIDA)}},
  title        = {AS Classification Dataset},
  howpublished = {\url{https://www.caida.org/catalog/datasets/as-classification/}},
  year         = {2015},
  note         = {Published September 10, 2015; last modified January 5, 2021; accessed May 15, 2025}
}

@inproceedings{borges:imc,
    author = {Carlos Selmo and Esteban Carisimo  and  Fabi\'an E. Bustamante and J. Ignacio Alvarez-Hamelin},
    title = {Learning AS-to-Organization Mappings with Borges},
    booktitle = imc,
    year = {2025}
}

@inproceedings{aleph:conext,
    author = {Kedar Thiagarajan and Esteban Carisimo  and  Fabi\'an E. Bustamante},
    title = {The Aleph: Decoding DNS PTR Records With Large Language Models},
    booktitle = conext,
    year = {2025}
}

@article{dainotti2016lost,
  title={Lost in space: improving inference of IPv4 address space utilization},
  author={Dainotti, Alberto and Benson, Karyn and King, Alistair and Huffaker, Bradley and Glatz, Eduard and Dimitropoulos, Xenofontas and Richter, Philipp and Finamore, Alessandro and Snoeren, Alex C},
  journal={IEEE Journal on Selected Areas in Communications},
  year={2016},
}

@misc{ripe2024personalasn,
  author       = {{RIPE Network Coordination Centre}},
  title        = {Personal AS Numbers, RIPE Open House},
  howpublished = {\url{https://www.ripe.net/meetings/open-house/personal-as-numbers-25-sept-2024/}},
  month        = sep,
  day          = {25},
  year         = {2024},
  note         = {Accessed May 16, 2025}
}

@misc{fiebig2024mauinmeowmeow,
  author       = {Tobias Fiebig},
  title        = {Putting the MAU Into meowmeow: On Personal ASNs},
  howpublished = {RIPE Labs Blog, \url{https://labs.ripe.net/author/tfiebig/putting-the-mau-into-meowmeow-on-personal-asns/}},
  month        = jul,
  day          = {22},
  year         = {2024},
  note         = {Accessed May 16, 2025}
}

@article{testart2024identifying,
  title={Identifying Current Barriers in RPKI Adoption},
  author={Testart, Cecilia and Wolff, Josephine and Gouda, Deepak and Fontugne, Romain},
  journal={Proc. of TPRC},
  year={2024}
}

@inproceedings{antonakakis2017understanding,
  title={Understanding the mirai botnet},
  author={Antonakakis, Manos and April, Tim and Bailey, Michael and Bernhard, Matt and Bursztein, Elie and Cochran, Jaime and Durumeric, Zakir and Halderman, J Alex and Invernizzi, Luca and Kallitsis, Michalis and others},
  booktitle=nsdi,
  year={2017}
}

@inproceedings{moore2002code,
  title={Code-Red: a case study on the spread and victims of an Internet worm},
  author={Moore, David and Shannon, Colleen and Claffy, K},
  booktitle=imc,
  year={2002}
}

@article{shin2011large,
  title={A large-scale empirical study of conficker},
  author={Shin, Seungwon and Gu, Guofei and Reddy, Narasimha and Lee, Christopher P},
  journal={IEEE ToIFS},
  year={2011}
}

@article{moore2003inside,
  title={Inside the slammer worm},
  author={Moore, David and Paxson, Vern and Savage, Stefan and Shannon, Colleen and Staniford, Stuart and Weaver, Nicholas},
  journal={Proc. of S\&P},
  year={2003}
}

@misc{ipinfo2025tags,
  author       = {{IPinfo}},
  title        = {Tags},
  howpublished = {\url{https://ipinfo.io/tags}},
  year         = {2025},
  note         = {Accessed May 18, 2025}
}

@inproceedings{carisimo:state_owned_ases,
  author = {Carisimo, Esteban and Gamero-Garrido, Alexander and Snoeren, Alex C. and Dainotti, Alberto},
  title     = {Identifying ASes of State-Owned Internet Operators},
  booktitle = imc,
  year      = {2021}
}

@inproceedings{kumar:govt_choices,
  title={Of Choices and Control - A Comparative Analysis of Government Hosting},
  author={Kumar, Rashna and Carisimo, Esteban and De Angelis Riva, Lukas and Buzzone, Mauricio and Bustamante, Fabi\'an E. and Qazi, Ihsan Ayyub and Beir\'o, Mariano G.},
  booktitle=imc,
  year={2024}
}

@inproceedings{richter2018advancing,
  title={Advancing the art of internet edge outage detection},
  author={Richter, Philipp and Padmanabhan, Ramakrishna and Spring, Neil and Berger, Arthur and Clark, David},
  booktitle=imc,
  year={2018}
}

@inproceedings{ager2011web,
  title={Web content cartography},
  author={Ager, Bernhard and M{\"u}hlbauer, Wolfgang and Smaragdakis, Georgios and Uhlig, Steve},
  booktitle=imc,
  year={2011}
}

@inproceedings{bajpai2017vantage,
  title={Vantage point selection for IPv6 measurements: Benefits and limitations of RIPE Atlas tags},
  author={Bajpai, Vaibhav and Eravuchira, Steffie Jacob and Sch{\"o}nw{\"a}lder, J{\"u}rgen and Kisteleki, Robert and Aben, Emile},
  booktitle={IFIP/IEEE Symposium on Integrated Network and Service Management (IM)},
  year={2017}
  }

@misc{bgptools2025tags,
  author       = {{BGP.Tools}},
  title        = {Tags},
  howpublished = {\url{https://bgp.tools/tags/}},
  year         = {2025},
  note         = {Accessed May 19, 2025}
}

@misc{nic2010pch,
  author       = {{NIC Chile}},
  title        = {New secondary service for .CL with Packet Clearing House},
  howpublished = {\url{https://www.nic.cl/anuncios/20100922-pch-eng.html}},
  month        = sep,
  day          = {22},
  year         = {2010},
  note         = {Santiago; accessed May 20, 2025}
}

@article{brown2020language,
  title   = {Language Models are Few-Shot Learners},
  author  = {Brown, Tom B. and Mann, Benjamin and Ryder, Nick and Subbiah, Melanie and Kaplan, Jared D. and Dhariwal, Prafulla and Neelakantan, Arvind and Shyam, Pranav and Sastry, Girish and Askell, Amanda and others},
  journal = {Proc. of NeurIPS},
  year    = {2020}
}

@inproceedings{richter2016multi,
  title={A multi-perspective analysis of carrier-grade NAT deployment},
  author={Richter, Philipp and Wohlfart, Florian and Vallina-Rodriguez, Narseo and Allman, Mark and Bush, Randy and Feldmann, Anja and Kreibich, Christian and Weaver, Nicholas and Paxson, Vern},
  booktitle=imc,
  year={2016}
}

@article{sherry2012making,
  title={Making middleboxes someone else's problem: Network processing as a cloud service},
  author={Sherry, Justine and Hasan, Shaddi and Scott, Colin and Krishnamurthy, Arvind and Ratnasamy, Sylvia and Sekar, Vyas},
  journal=sigcomm,
  year={2012}
}

@article{lodhi2014using,
  title={Using peeringDB to understand the peering ecosystem},
  author={Lodhi, Aemen and Larson, Natalie and Dhamdhere, Amogh and Dovrolis, Constantine and Claffy, Kc},
  journal=sigcomm-ccr,
  year={2014},
}

@misc{microsoft:peering:policy,
  title =  {Prerequisites to set up peering with {Microsoft}},
  author =   {Microsoft},
  howpublished =
                  {\url{https://docs.microsoft.com/en-us/azure/internet-peering/prerequisites}},
  year =   2022,
}

@misc{google:peering:policy,
  title =  {Prerequisites to Peer with {Google}},
  author =   {Google},
  howpublished = {\url{https://peering.google.com/\#/options/peering}},
  year =   2022,
}

@misc{facebook:peering:policy,
  title =  {Peering: Technical Requirements},
  author =   {Facebook},
  howpublished = {\url{https://www.facebook.com/peering}},
  year =   2022,
}

@misc{eyeballs1,
  author="Geoff Huston",
  title="{How Big is that Network?}",
  howpublished="\url{https://labs.apnic.net/?p=526}",
    year={2014},
}

@inproceedings{apnic:imc24,
author = {Salamatian, Loqman and Ardi, Calvin and Giotsas, Vasileios and Calder, Matt and Katz-Bassett, Ethan and Arnold, Todd},
title = {What's in the Dataset? Unboxing the APNIC per AS User Population Dataset},
year = {2024},
booktitle = imc
}

@misc{apnic_eyeballs,
    author = "{{APNIC}}",
    title = {{Customers per AS Measurements}},
    howpublished={\url{https://stats.labs.apnic.net/aspop/}},
    year={2020},
}

@inproceedings{labovitz2010internet,
  title =  {Internet inter-domain traffic},
  author =   {Labovitz, Craig and Iekel-Johnson, Scott and
                  McPherson, Danny and Oberheide, Jon and Jahanian,
                  Farnam},
  booktitle =  sigcomm,
  year =   2010
}

@article{yook2002modeling,
  title={Modeling the Internet's large-scale topology},
  author={Yook, Soon-Hyung and Jeong, Hawoong and Barab{\'a}si, Albert-L{\'a}szl{\'o}},
  journal={PNAS},
  year={2002}
}

@incollection{sechidis2011stratification,
  title        = {On the stratification of multi-label data},
  author       = {Sechidis, Konstantinos and Tsoumakas, Grigorios and Vlahavas, Ioannis},
  booktitle    = {Machine Learning and Knowledge Discovery in Databases. ECML PKDD 2011},
  year         = {2011}
}

@inproceedings{szymanski2017network,
  title        = {A Network Perspective on Stratification of Multi-Label Data},
  author       = {Szyma{\'n}ski, Piotr and Kajdanowicz, Tomasz},
  booktitle    = {Proc of Workshop on Learning with Imbalanced Domains: Theory and Applications},
  year         = {2017}
}

@misc{caida2025,
  author       = {{Center for Applied Internet Data Analysis (CAIDA)}},
  title        = {CAIDA},
  howpublished = {\url{https://www.caida.org}},
  year         = {2025},
  note         = {Accessed June 2, 2025}
}

@misc{peeringdb2025,
  author       = {PeeringDB},
  title        = {PeeringDB},
  howpublished = {\url{https://www.peeringdb.com}},
  year         = {2025},
  note         = {Accessed June 2, 2025}
}

@misc{naics2025,
  author       = {{Office of Management and Budget (OMB)}},
  title        = {North American Industry Classification System (NAICS)},
  howpublished = {\url{https://www.census.gov/naics/}},
  year         = {2025},
  note         = {Accessed June 2, 2025}
}

@article{openai2023gpt4,
  author       = {{OpenAI}},
  title        = {GPT-4 Technical Report},
  journal      = {OpenAI},
  year         = {2023},
  url          = {https://openai.com/research/gpt-4},
  note         = {Accessed June 2, 2025}
}

@article{touvron2023llama,
  author       = {Touvron, Hugo and Martin, Thibault and Stone, Chloe and Albert, Zhiqing and Almahairi, Amjad and Babaei, Armand and Bloom, Jacob and Boecking, Benjamin and Cai, Liv and Caron, M\'elanie and Douze, Matthijs and Erichson, Niels and Fong, Robert and Gallangher, John and Ghadiyaram, Deepti and Grill, Jean-Bastien and Gu, Lucy and Henderson, Jonah and Hernandez, Daniel and Hogwild, Michael and Joglekar, Shruti and Khlaaf, Hatem and Klein, Guillaume and Li, Xiang Lisa and Livi, Lorenzo and Ly, Conrad and Maillard, Jean-Bastien and Miao, Yi and Milan, Sophie and Misra, Ishan and Nugteren, Jaap and Ortega, Paco and Pai, Archit and Pezeshki, Mehrad and Ramesh, Aditya and Ramesh, Karthik and Rigotti, Stefano and Robertson, Zoe and Rogozhnikov, Alexey and Roufosse, Charles and Sandler, Mark and Smola, Alex and Sundar, Pratik and Sewerin, Philip and Tapia, Enrique and Touvron, Hugo and Turner, Cole and Uzuner, Ozan and Wang, Reid and Yaffe, Victoria and Zellers, Rowan and Zhang, Jason and Zolfaghari, Mohammad},
  title        = {{LLaMA: Open and Efficient Foundation Language Models}},
  journal      = {arXiv preprint arXiv:2302.13971},
  year         = {2023}
}

@misc{caida2025asrank,
  author       = {{Center for Applied Internet Data Analysis (CAIDA)}},
  title        = {AS Rank},
  howpublished = {\url{https://asrank.caida.org}},
  year         = {2025},
  note         = {Accessed June 4, 2025}
}

@misc{ipinfo2025,
  author       = {{IPinfo}},
  title        = {IPinfo},
  howpublished = {\url{https://ipinfo.io}},
  year         = {2025},
  note         = {Accessed June 4, 2025}
}

@misc{ietf2015rdap,
  author       = {{IETF RDAP Working Group}},
  title        = {Registration Data Access Protocol (RDAP) Query Format},
  howpublished = {RFC 7480, \url{https://tools.ietf.org/html/rfc7480}},
  month        = mar,
  year         = {2015},
  note         = {Accessed June 4, 2025}
}

@inproceedings{nemmi2021parallel,
  title={The parallel lives of autonomous systems: ASN allocations vs. BGP},
  author={Nemmi, Eugenio Nerio and Sassi, Francesco and La Morgia, Massimo and Testart, Cecilia and Mei, Alessandro and Dainotti, Alberto},
  booktitle=imc,
  year={2021}
}

@misc{ripe2025quarantine,
  author       = {{RIPE Network Coordination Centre}},
  title        = {Quarantine for Returned Internet Number Resources},
  howpublished = {\url{https://www.ripe.net/manage-ips-and-asns/resource-management/quarantine-for-returned-internet-number-resources/}},
  year         = {2025},
  note         = {Accessed June 4, 2025}
}

@misc{arin2018fees,
  author       = {{American Registry for Internet Numbers (ARIN)}},
  title        = {2018 Fee Schedule},
  howpublished = {\url{https://www.arin.net/resources/fees/fee_schedule/2018_fee_schedule/}},
  year         = {2018},
  note         = {Accessed June 4, 2025}
}

@misc{anghel2024driving,
  author       = {Anghel, Radu},
  title        = {Driving the ASN Truck Without a Licence},
  howpublished = {\url{https://labs.ripe.net/author/eu/driving-the-asn-truck-without-a-licence/}},
  month        = jul,
  day          = {18},
  year         = {2024},
  note         = {Accessed June 4, 2025}
}

@misc{isoc2023languages,
  author       = {{Internet Society Foundation}},
  title        = {What are the most used languages on the Internet?},
  howpublished = {\url{https://www.isocfoundation.org/2023/05/what-are-the-most-used-languages-on-the-internet/}},
  month        = may,
  day          = {15},
  year         = {2023},
  note         = {Accessed June 5, 2025}
}

@misc{sklearn_precision_score_docs,
  author       = {{scikit-learn developers}},
  title        = {scikit-learn: \texttt{sklearn.metrics.precision\_score} documentation},
  year         = {2025},
  howpublished = {\url{https://scikit-learn.org/stable/modules/generated/sklearn.metrics.precision_score.html}},
  note         = {Accessed: 2025-10-30}
}

@article{xu2024hallucination,
  title={Hallucination is inevitable: An innate limitation of large language models},
  author={Xu, Ziwei and Jain, Sanjay and Kankanhalli, Mohan},
  journal={arXiv preprint arXiv:2401.11817},
  year={2024}
}

@article{chae2025large,
  title={Large language models for text classification: from zero-shot learning to instruction-tuning},
  author={Chae, Youngjin and Davidson, Thomas},
  journal={Sociological Methods \& Research},
  pages={00491241251325243},
  year={2025},
  publisher={SAGE Publications Sage CA: Los Angeles, CA}
}

@article{divya2024comparing,
  title={Comparing human text classification performance and explainability with large language and machine learning models using eye-tracking},
  author={Divya Venkatesh, Jeevithashree and Jaiswal, Aparajita and Nanda, Gaurav},
  journal={Scientific reports},
  year={2024}
}

\appendix
\section{Detailed Category Definitions}\label{sec:appendix:taxonomy_details}

This appendix provides detailed definitions, sub-tier descriptions, and representative examples for each of \name{}'s 18 top-level categories.

\noindent{\textbf{Access Networks.}} \textit{Small access networks} serve a limited number of subscribers --- typically a few thousand --- and cover narrow geographic areas, such as a neighborhood or small town. They operate with lower capacity and simpler infrastructure. Examples include home or SOHO fiber providers (e.g., Atherton Fiber, AS53859 in California), community wireless meshes (e.g., Ninux.org, AS197835 in Italy), and small-scale or boutique ISPs (e.g., Guifi.net, AS49835 in Spain; XMission, AS6315 in Utah). \textit{Large access networks} serve millions of subscribers and often span entire countries. Due to ongoing market consolidation, most populous nations have at least one such dominant provider. Notable examples include Jio and Airtel in India, China Telecom and China Mobile, and AT\&T and Comcast in the United States.

\noindent{\textbf{Transit Providers.}} \textit{Domestic transits} primarily serve downstream networks within a single country; their interconnection footprint is limited to local IXPs and domestic peering points. \textit{Regional transits} span multiple countries within the same continent, with customer cones rarely extending beyond it. \textit{Global transits} operate at a worldwide scale, with infrastructure crossing continental boundaries (e.g., Lumen, AS3356).

\noindent{\textbf{Non-Terrestrial Networks.}} \textit{Satellite operators} include networks with access to GEO, MEO, and LEO constellations. \textit{Mobile providers} include cellular operators with their own ASNs.

\noindent{\textbf{DNS Infrastructure.}} \textit{Root DNS ASes} represent the 12 organizations hosting the 13 root letters. \textit{(cc)TLD registry ASes} correspond to dedicated ASNs that run individual country-code or generic TLDs (e.g., DENIC operates ``.de'' on AS8763; AFNIC serves ``.fr'' on AS2484/AS2485/AS2486). Some registries outsource part of their operation (e.g., Chile's ``.cl'' relies on PCH as a secondary provider~\cite{nic2010pch}). \textit{Authoritative DNS hosting ASes} provide managed DNS services at scale (e.g., GoDaddy).

\noindent{\textbf{Content Providers.}} \textit{CDN} includes ASes that distribute cached content via widespread edge footprints (e.g., Akamai, Fastly). \textit{Cloud} includes elastic computing platforms that expose on-demand virtualized resources, autoscaling, and programmable APIs (e.g., AWS, GCP, Azure). \textit{Hosting} includes traditional shared, VPS, or dedicated-server providers lacking the elastic orchestration characteristics of clouds. Providers self-advertising as ``cloud'' but offering only static hosting plans are placed in this last category.

\noindent{\textbf{Educational and Research.}} \textit{Universities} (e.g., MIT's AS3), \textit{schools} (e.g., Pasadena Independent School District's AS63171), \textit{research institutes} (e.g., Lawrence Livermore National Laboratory's AS44), and \textit{academic backbones} (e.g., RedClara's AS27750), which serve as primary Internet providers for multiple educational organizations.

\noindent{\textbf{Government.}} \textit{Jurisdiction} specifies the level of government responsible for the AS: \textit{national}, \textit{state/province}, or \textit{city/county/municipality}. \textit{Branch} identifies the branch of government overseeing the network: \textit{executive}, \textit{legislative}, \textit{judicial}, or \textit{none}. \textit{Role} includes three non-exclusive Boolean tags indicating whether the AS acts as a \textit{regulator} (e.g., Brazil's telecom regulator Anatel, AS61656), a \textit{government agency} (e.g., NASA, AS24), or serves other functions.

\noindent{\textbf{Law Enforcement.}} Includes police departments (e.g., United States Capitol Police, AS63147), coast-guard services (e.g., Bangladesh Coast Guard Force, AS137415), and other national or sub-national law-enforcement bodies.

\noindent{\textbf{Enterprise.}} \textit{Technology} (e.g., Intel Corporation, AS4983), \textit{manufacturing/industrial} (e.g., Siemens, AS8826), \textit{e-commerce} (e.g., eBay, AS62955), and \textit{entertainment} (e.g., AMC Theatres, AS14509).

\noindent{\textbf{Energy \& Utilities.}} \textit{Electricity} (e.g., National Grid, AS11529), \textit{gas} (e.g., New Mexico Gas Company, AS399479), \textit{water} (e.g., Veolia Eau, AS209513), and \textit{energy/oil} (e.g., ExxonMobil, AS1766).

\noindent{\textbf{Finance.}} \textit{Commercial banks} (e.g., Bank of America, AS10794), \textit{central banks} (e.g., Federal Reserve Bank, AS7466), \textit{credit unions} (e.g., Navy Federal Credit Union, AS14222), and \textit{stock exchanges} (e.g., NASDAQ, AS54811).

\noindent{\textbf{Health.}} \textit{Hospitals} (e.g., Cleveland Clinic, AS22093) and \textit{insurance} providers (e.g., UnitedHealth Group, AS10879).

\noindent{\textbf{TV, Radio \& Cultural Amenities.}} \textit{TV channels} (e.g., BBC, AS2818), \textit{radio stations} (e.g., NPR, AS25755), and \textit{libraries/museums} (e.g., National Library of China, AS9401). Although many Internet providers are affiliated with cable companies that own TV channels or radio stations, this category exclusively considers networks dedicated to a single channel or station.

\noindent{\textbf{Transportation.}} \textit{Airlines/airports} (e.g., Emirates Airline, AS6168), \textit{railways} (e.g., France's SNCF, AS203885), \textit{shipping companies} (e.g., Royal Caribbean, AS35955), and \textit{transit authorities} (e.g., New York MTA, AS12187).

\section{Mobile Virtual Network Operators}\label{sec:appendix:mvno}

Table~\ref{tab:mvno-asn} displays the list of Autonomous Systems operated by Mobile Virtual Network Operators (MVNOs).

\begin{table*}[htbp]
  \centering
  \footnotesize
  \begin{tabular}{llp{10cm}}
    \toprule
    \textbf{MVNO} & \textbf{ASN} & \textbf{Description} \\
    \midrule
    Truphone                          & AS30967  & Global eSIM/M2M operator headquartered in the UK; peers on several IXPs worldwide. \\[2pt]
    Lebara Mobile                     & AS33921  & UK-based ethnic/low-cost MVNO active across Europe. \\[2pt]
    Lycamobile USA                    & AS23275  & U.S. subsidiary of the multinational Lycamobile group. \\[2pt]
    Lycamobile (CEE region)           & AS204859 & Smaller Lycamobile entity registered in Central/Eastern Europe. \\[2pt]
    Vectone Mobile / Mundio Mobile    & AS39477  & Budget MVNO operating in the UK and parts of Europe under several brands (Vectone, Delight, etc.). \\[2pt]
    Boost Mobile (US)                 & AS398378 & After the 2020 Dish purchase, Boost began announcing its own prefixes for VoIP/SMS back-end services. \\[2pt]
    Tracfone Wireless                 & AS32091  & Legacy Tracfone core network (now owned by Verizon). \\[2pt]
    Consumer Cellular                 & AS396875 & Senior-focused MVNO in the U.S.; runs a small BGP footprint for customer-facing services. \\
    \bottomrule
  \end{tabular}
  \caption{Example Mobile Virtual Network Operators (MVNOs) and their Autonomous System Numbers (ASNs).}
  \label{tab:mvno-asn}
\end{table*}

\section{Annotated Dataset}\label{sec:appendix:annotated}

In this appendix, we detail the gold-standard corpus, presenting per-class statistics (App.~\ref{sec:appendix:annotated:class_stats}) and feature availability for each class (App.~\ref{sec:appendix:annotated:availability}).

\subsection{Statistics of the Annotated Dataset}\label{sec:appendix:annotated:class_stats}

\begin{table*}[htbp]
  \footnotesize
    \centering
    \setlength{\tabcolsep}{3pt}
    \caption{Distribution of annotated AS records by category and subcategory.}
    \label{tab:annotated_tags}
    \begin{minipage}[t]{0.31\textwidth} 
    \vspace{0pt} 
    \begin{tabular}{m{3.3cm} r}
      \toprule
      \textbf{Category/Subcategory} & \textbf{Count} \\
      \midrule
      \textbf{Access} & \textbf{315} \\
      \quad Large ISP & 92 \\
      \quad Small ISP & 223 \\
      \midrule
      \textbf{Transit} & \textbf{98} \\
      \quad Global & 38 \\
      \quad Regional & 33 \\
      \quad Domestic & 27 \\
      \midrule
      \textbf{Mobile} & \textbf{103} \\
      \quad Mobile & 103 \\
      \midrule
      \textbf{Satellite} & \textbf{41} \\ 
      \quad Satellite & 41 \\
      \midrule
      \textbf{Government} & \textbf{1192} \\ 
      \quad Executive & 396 \\
      \quad Legislative & 28 \\
      \quad Judiciary & 64 \\
      \quad National & 394 \\
      \quad State/Province & 105 \\
      \quad City/County/Muni. & 67 \\
      \quad Regulators & 33 \\
      \quad Agencies & 105 \\
      \midrule
      \textbf{IXP} & \textbf{145} \\ 
      \quad IXP & 145 \\
      \bottomrule
    \end{tabular}
    \end{minipage}
    \hfill
    \begin{minipage}[t]{0.31\textwidth} 
    \vspace{0pt} 
    \begin{tabular}{m{3.3cm} r}
      \toprule
      \textbf{Category/Subcategory} & \textbf{Count} \\
      \midrule
      \textbf{Content Provider} & \textbf{184} \\ 
      \quad Cloud & 33 \\
      \quad Hosting & 113 \\
      \quad CDN & 38 \\
      \midrule
      \textbf{DNS} & \textbf{144} \\ 
      \quad Roots & 84 \\
      \quad ccTLD & 32 \\
      \quad ANS & 28 \\
      \midrule
      \textbf{Law Enforcement} & \textbf{54} \\ 
      \quad Law Enforcement & 54 \\
      \midrule
      \textbf{Enterprise} & \textbf{195} \\ 
      \quad E-commerce & 25 \\
      \quad Entertainment & 33 \\
      \quad Industrial Manufact. & 22 \\
      \quad Technology & 115 \\
      \midrule
      \textbf{Cooperatives} & \textbf{43} \\ 
      \quad Cooperatives & 43 \\
      \midrule
      \textbf{Transportation} & \textbf{111} \\ 
      \quad Trains & 21 \\
      \quad Ships & 24 \\
      \quad Airports & 44 \\
      \quad Transit Authority & 22 \\
      \bottomrule
    \end{tabular}
    \end{minipage}
    \hfill
    \begin{minipage}[t]{0.31\textwidth} 
    \vspace{0pt} 
    \begin{tabular}{m{3.3cm} r}
      \toprule
      \textbf{Category/Subcategory} & \textbf{Count} \\
      \midrule
      \textbf{Financial} & \textbf{205} \\ 
      \quad Bank & 91 \\
      \quad Central Banks & 25 \\
      \quad Credit Union & 23 \\
      \quad Entities & 33 \\
      \quad Stock Exchanges & 33 \\
      \midrule
      \textbf{Health} & \textbf{53} \\ 
      \quad Insurances & 26 \\
      \quad Hospitals & 27 \\
      \midrule
      \textbf{Educational \& Research} & \textbf{237} \\ 
      \quad University & 59 \\
      \quad Academic Backbone & 32 \\
      \quad Schools & 80 \\
      \quad Research Institutes & 66 \\
      \midrule
      \textbf{TV/Radio/Cultural} & \textbf{40} \\ 
      \quad TV/Radio/Cultural & 40 \\
      \midrule
      \textbf{Energy \& Utility} & \textbf{63} \\ 
      \quad Energy \& Utility & 63 \\
      \midrule
      \textbf{Personal} & \textbf{31} \\ 
      \quad Personal & 31 \\
      \bottomrule
    \end{tabular}
    \end{minipage}
\end{table*}

Table~\ref{tab:annotated_tags} summarizes the number of ASes in each top-level and sub-level class.

\subsection{Data Availability}\label{sec:appendix:annotated:availability}
\begin{figure*}[htbp]
  \centering
  \includegraphics[width=\textwidth]{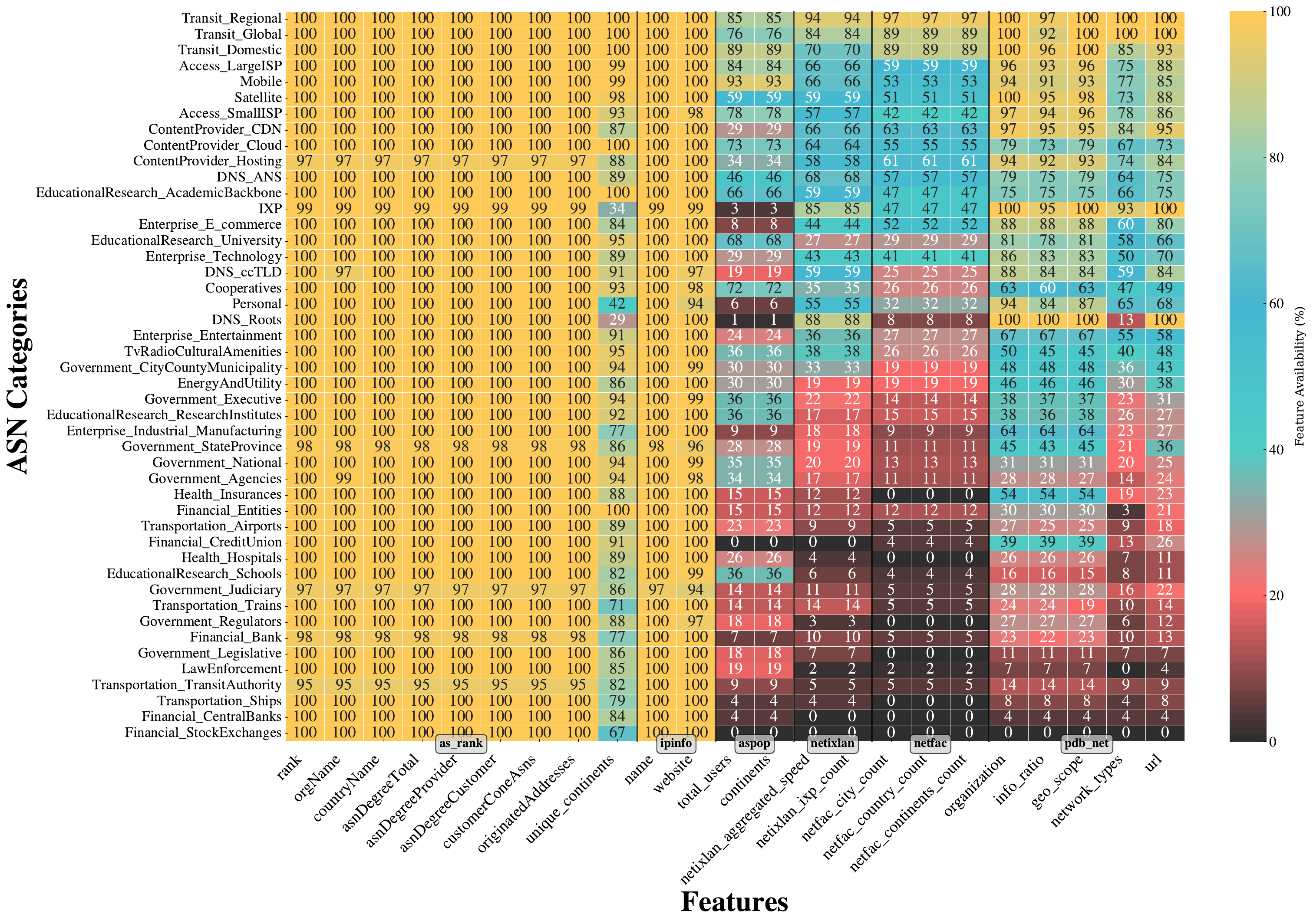}
  \caption{Relative availability of each feature for the annotated ASes in our dataset, grouped by category.
  }
  \label{fig:heatmap}
\end{figure*}

Figure~\ref{fig:heatmap} shows a heatmap that displays, for each class in the gold-standard corpus, the percentage of networks for which each feature is available.

\section{\name{} trained with other taxonomies}

\subsection{\name{} trained with ASdb labels}\label{sec:appendix:asdb_training}

Table~\ref{tab:dry_run} presents the accuracy, precision, and recall for \name{} when it is trained with ASdb labels.

\begin{table*}[ht]
  \scriptsize
  \centering
  \setlength{\tabcolsep}{2.5pt}
  \caption{Per-tag metrics (Prec: Precision, Rec: Recall, Acc: Accuracy) for the \name{} model (fine-tuned) to the ASdb (NAICSlite) taxonomy using the predictions from the ASdb model as Ground Truth. Average per-label accuracy = 0.95, Macro Precision = 0.57, Macro Recall = 0.48, Macro F1 = 0.50, Overall Accuracy = 0.55.}
  \label{tab:dry_run}
  \begin{tabular*}{\textwidth}{@{\extracolsep{\fill}} l c c c  l c c c @{}}
    \toprule
    \textbf{Tag} & \textbf{Prec.} & \textbf{Rec.} & \textbf{Acc.}
                & \textbf{Tag} & \textbf{Prec.} & \textbf{Rec.} & \textbf{Acc.} \\
    \midrule
    \textbf{Other}                         & --   & --   & --   
                                   & \textbf{Retail \& E-commerce}        & 0.55 & 0.18 & 0.94 \\
    \textbf{Agriculture, Mining \& Refining} & 0.25 & 0.27 & 0.97 
                                   & \textbf{Freight \& Postal}           & 0.54 & 0.61 & 0.96 \\
    \textbf{IT \& Software}                & 0.83 & 0.87 & 0.83 
                                   & \textbf{Finance \& Insurance}        & 0.83 & 0.82 & 0.96 \\
    \textbf{Education \& Research}         & 0.75 & 0.81 & 0.95 
                                   & \textbf{Travel \& Lodging}           & 0.63 & 0.67 & 0.98 \\
    \textbf{Manufacturing}                 & 0.35 & 0.26 & 0.95 
                                   & \textbf{Government \& Public Admin}  & 0.64 & 0.69 & 0.90 \\
    \textbf{NGOs \& Community}             & 0.77 & 0.46 & 0.94 
                                   & \textbf{Media \& Publishing}         & 0.74 & 0.56 & 0.94 \\
    \textbf{Health Care}                   & 0.61 & 0.69 & 0.98 
                                   & \textbf{Utilities (non-ISP)}         & 0.40 & 0.40 & 0.97 \\
    \textbf{Museums \& Entertainment}      & 0.20 & 0.11 & 0.98
                                   & \textbf{Services (General)}          & 0.54 & 0.21 & 0.89 \\
    \bottomrule
  \end{tabular*}
\end{table*}

\subsection{\name{}'s Performance with UN's ISIC taxonomy}\label{sec:appendix:isic}

Table~\ref{tab:isic_results} shows the per-class Accuracy, Precision, and Recall of \name{} using the UN's ISIC taxonomy.

\begin{table*}[ht]
  \scriptsize
  \centering
  \setlength{\tabcolsep}{2.5pt}
  \caption{Per-tag metrics (Prec: Precision, Rec: Recall, Acc: Accuracy) for the \name{} model (fine-tuned) to the ISIC taxonomy.
  Average per-label accuracy = 0.99, Macro Precision = 0.85, Macro Recall = 0.84, Macro F1 = 0.83, Exact-Match Accuracy = 0.83.}
  \label{tab:isic_results}
  \begin{tabular*}{\textwidth}{@{\extracolsep{\fill}} l c c c  l c c c @{}}
    \toprule
    \textbf{Tag} & \textbf{Prec.} & \textbf{Rec.} & \textbf{Acc.}
                & \textbf{Tag} & \textbf{Prec.} & \textbf{Rec.} & \textbf{Acc.} \\
    \midrule
  	\textbf{Construction}                         & 1.00 & 1.00 & 1.00  
                  & \textbf{Arts, Entertain. \& Recreation}           & 0.67 & 0.67 & 1.00 \\
  	\textbf{Real Estate Act} & 1.00 & 1.00 & 1.00 
                   & \textbf{Transport \& Storage}           & 0.90 & 0.87 & 0.99 \\
  	\textbf{Agriculture, Forestry \& Fishing}                & 0.83 & 0.83 & 1.00 
                   & \textbf{Financial \& Insurance Act}        & 0.91 & 0.92 & 0.98 \\
  	\textbf{Administrative \& Support Serv}         & 0.75 & 0.75 & 1.00 
                  & \textbf{Education}        & 0.91 & 0.89 & 0.99 \\
  	\textbf{Other Service Act}                 & 1.00 & 0.50 & 1.00 
                   & \textbf{Accommodation \& Food Serv}  & 0.80 & 1.00 & 1.00 \\
  	\textbf{Extraterritorial Organizations}             & 0.67 & 0.67 & 1.00 
                   & \textbf{Mining \& Quarrying}         & 0.60 & 1.00 & 1.00 \\
  	\textbf{Professional, Science \& Technical}                   & 0.59 & 0.86 & 0.97 
                   & \textbf{Household Activities}         & 1.00 & 0.78 & 1.00 \\
  	\textbf{Manufacturing}      & 0.62 & 0.80 & 0.99 
                   & \textbf{Trade \& Repair of Vehicles}          & 0.88 & 1.00 & 1.00 \\
  	\textbf{Public Admin \& Defense}      & 0.95 & 0.77 & 0.93 
                   & \textbf{Human Health \& Social Work}          & 0.93 & 0.72 & 0.99 \\
  	\textbf{Information \& Communication}      & 0.97 & 0.93 & 0.95
                   & \textbf{Utilities \& Waste Mgmt}          & 0.94 & 0.89 & 1.00 \\
    \bottomrule
  \end{tabular*}
\end{table*}

\section*{Ethics}\label{sec:ethics}

This work raises no ethical concerns. Our study is based exclusively on publicly available datasets, including PeeringDB~\cite{peeringdb2025}, CAIDA AS relationship data~\cite{caida2025}, BGP routing tables from RouteViews and RIPE RIS, IPinfo's AS metadata, and RDAP registry records. No human subjects were recruited or studied, and no personally identifiable information was collected or processed. The annotated dataset of 2,000 ASes records organizational-level labels (network type, sector, jurisdiction) and does not contain individual-level data. We will release the annotated dataset and classification code to support reproducibility and further research.


\end{document}